\documentclass[twocolumn,superscriptaddress,reprint,nofootinbib,prd]{revtex4-1}%
\usepackage{amssymb}
\usepackage{color}
\usepackage{graphicx}
\usepackage{multirow}
\usepackage{dcolumn}
\usepackage{bm}
\usepackage[header,title,page,titletoc]{appendix}
\usepackage{amsmath}
\usepackage{bbm}
\usepackage{amsfonts}%
\usepackage[bookmarks,colorlinks=false]{hyperref}
\usepackage{ulem}
\newlength{\wordlength}

\newlength{\onewordlength}

    \newcommand{\ba}{\begin{eqnarray}}
    \newcommand{\ea}{\end{eqnarray}}
    \newcommand{\be}{\begin{equation}}
    \newcommand{\ee}{\end{equation}}

\newcommand{\bzero}{{\bf 0}}
\newcommand {\btheta} {{\mbox{\boldmath$\theta$}}}

\newcommand{\bfe}{{\bf e}}
\newcommand{\bn}{{\bf n}}

\newcommand {\bk} {{\mathbf k}}

\newcommand {\bp} {{\mathbf p}}

\newcommand{\bx}{{\bf x}}

\newcommand{\calP}{{\mathcal P}}

\newcommand{\calO}{{\mathcal O}}
\newcommand{\calR}{{\mathcal R}}

\newcommand{\calZ}{{\mathcal Z}}

\newcommand{\calV}{{\mathcal V}}

\begin{document}
%\runauthor{PKU}
%\begin{frontmatter}

\title{Low-energy Scattering of $(D\bar{D}^{*})^\pm$ System And the Resonance-like Structure $Z_c(3900)$}

\author{Ying Chen}
\affiliation{%
Institute of High Energy Physics, Chinese Academy of Sciences, Beijing 100049, China
}%

\author{Ming Gong}
\affiliation{%
Institute of High Energy Physics, Chinese Academy of Sciences, Beijing 100049, China
}
\author{Yu-Hong Lei}
\affiliation{%
School of Physics, Peking University, Beijing 100871, China
}%

\author{Ning Li}
\affiliation{%
School of Physics, Peking University, Beijing 100871, China
}%

\author{Jian Liang}
\affiliation{%
Institute of High Energy Physics, Chinese Academy of Sciences, Beijing 100049, China
}%

\author{Chuan Liu}%
\email[Corresponding author. Email: ]{liuchuan@pku.edu.cn}
\affiliation{%
School of Physics and Center for High Energy Physics, Peking
University, Beijing 100871, China
}%

\author{Hang Liu}
\affiliation{%
School of Physics, Peking University, Beijing 100871, China
}

\author{Jin-Long Liu}
\affiliation{%
School of Physics, Peking University, Beijing 100871, China
}

\author{Liuming Liu}
\affiliation{%
Helmholtz-Institut f\"ur Strahlen-und Kernphysik and Bethe Center for Theoretical Physics, Universit\"at Bonn, D-53115 Bonn, Germany
}

\author{Yong-Fu Liu}
\affiliation{%
School of Physics, Peking University, Beijing 100871, China
}

\author{Yu-Bin Liu}
\affiliation{%
School of Physics, Nankai University, Tianjin 300071, China
}

\author{Zhaofeng Liu}
\affiliation{%
Institute of High Energy Physics, Chinese Academy of Sciences, Beijing 100049, China
}

\author{Jian-Ping Ma}
\affiliation{%
Institute of Theoretical Physics, Chinese Academy of Sciences, Beijing 100190, China
}

\author{Zhan-Lin Wang}
\affiliation{%
School of Physics, Peking University, Beijing 100871, China
}

\author{Yi-Bo~Yang}
\affiliation{%
Institute of High Energy Physics, Chinese Academy of Sciences, Beijing 100049, China
}

\author{Jian-Bo~Zhang}
\affiliation{%
Department of Physics, Zhejiang University, Hangzhou 311027, China
}
 \begin{center}
 (CLQCD Collaboration)
 \end{center}
 \begin{abstract}
 In this exploratory lattice study, low-energy scattering of the $(D\bar{D}^{*})^\pm$ meson system is analyzed using lattice QCD with $N_f=2$ twisted mass fermion configurations with three pion mass values.
 The calculation is performed within single-channel L\"uscher's finite-size
 formalism. The threshold scattering parameters, namely the scattering
 length $a_0$ and the effective range $r_0$, for the $s$-wave scattering in
 $J^P=1^+$ channel are extracted. For the cases in our study, the interaction
 between the two charmed mesons is weakly repulsive. Our lattice results therefore
 do not support  the possibility of a shallow bound state
 for the two mesons for the pion mass values we studied. This calculation
 provides some useful information on the nature of the newly
 discovered resonance-like structure $Z_c(3900)$ by various experimental groups.
 \end{abstract}

 \maketitle

% \begin{widetext}

% \end{widetext}
%\newpage

 \section{Introduction}

 Recently, a charged resonance-like structure $Z^{\pm}_c(3900)$ has been observed at
 BESIII in the $\pi^\pm J/\psi$ invariant mass spectrum from the $Y(4260)$ decays~\cite{Ablikim:2013mio}.
 The same structure was confirmed shortly by the Belle~\cite{Liu:2013dau} and CLEO collaborations~\cite{Xiao:2013iha}.
 This discovery has triggered many theoretical investigations on the nature of this
 structure, see e.g. Ref.~\cite{Wang:2013cya} and references therein.
 It is readily observed that the invariant mass of the structure is close to the $DD^*$ threshold,
 one possible interpretation is a molecular bound state formed by the $\bar{D}^*$ and $D$ mesons.
 Other possibilities have also been discussed.
 To further investigate these possibilities, the interaction between
 $\bar{D}^*$ and $D$ mesons (or the conjugated systems under $C$-parity or $G$-parity, e.g.
 $\bar{D}^0D^{*\pm}$, $D^\pm\bar{D}^{*0}$, etc.) becomes important.
 All these possible meson systems will be generically denoted as $(D\bar{D}^*)^\pm$ systems
 in what follows.
 As is known, the interaction of two hadrons can be studied via
 the scattering process of the relevant hadrons. Since the energy being
 considered here is very close to the threshold of the $(D\bar{D}^{*})^\pm$ system,
 only threshold scattering parameters, i.e. scattering length $a_0$ and
 effective range $r_0$, are relevant for this particular study.
 In phenomenological studies, the interaction between the mesons can be computed
 by assuming meson exchanges models.
 However, since the interaction between the charmed mesons at low-energies
 is non-perturbative in nature, it is tempting
 to study the problem using a genuine non-perturbative method like lattice QCD.

 In this paper, we study the scattering threshold parameters of
 $(D\bar{D}^{*})^\pm$ system using lattice QCD within
 the single-channel L\"uscher's formalism, a finite-size technique
 developed to study scattering processes in a finite
 volume~\cite{luscher86:finitea,luscher86:finiteb,luscher90:finite,luscher91:finitea,luscher91:finiteb}.
 In this exploratory study, $N_f=2$ twisted mass gauge field configurations are utilized.
 Since the binding (or unbinding) nature of the state can depend sensitively on
 the value of the pion mass, as is the case for baryon-baryon systems,
 we have utilized three different values of pion mass
 corresponding to $m_\pi=485, 420, 300 MeV$, respectively, allowing us to
 investigate the pion mass dependence of our results.
 The size of the lattices is $32^3\times 64$ with a lattice spacing of
 about $0.067fm$. The computation is carried out in the $J^P=1^+$ channel. We find that,
 in this particular channel, the interaction between the two constituent mesons
 is weakly repulsive in nature and our results therefore do not support a bound state
 of the two mesons. This is in agreement with a similar recent lattice study using two flavor improved Wilson fermions~\cite{Prelovsek:2013xba,Prelovsek:2013sxa}, which is carried out with one pion mass value
 and a smaller lattice. In a different channel ($J^{PC}=1^{++}$),
 the authors of the previous references have also found interesting evidence for
 the puzzling $X(3872)$~\cite{Prelovsek:2013cra}.

 This paper is organized as follows. In Section~\ref{sec:method}, we briefly
 introduce L\"uscher's formalism. In Section~\ref{sec:operators}, one-particle
 and two-particle interpolating operators and their correlation matrices
 are defined. In section~\ref{sec:simulation_details}, simulation
 details are given and the results for the single- and two-meson systems
 are analyzed.  By applying L\"uscher's
 formula the scattering phases are extracted for various lattice momenta.
 When fitted to the known low-energy behavior, the threshold parameters
 of the system, i.e. the inverse scattering length $a^{-1}_0$ and the effective range $r_0$
 are obtained. We also discuss possible multi-channel effects that might affect our results.
 In Section~\ref{sec:conclude}, we will conclude with some general remarks.

 \section{Strategies for the computation}
 \label{sec:method}

% \subsection{Twisted boundary conditions}

 Within L\"uscher's formalism, the exact energy eigenvalue
 of a two-particle system in a finite box of size $L$ is
 related to the elastic scattering phase of the two particles
 in the infinite volume. Consider two interacting particles
 with mass $m_1$ and $m_2$ enclosed in a cubic box of size $L$,
 with periodic boundary conditions applied in all three directions.
 The spatial momentum $\bk$ is quantized according to:
 \be
 \label{eq:free_k}
 \bk=\left({2\pi\over L}\right)\bn\;,
 \ee
 with $\bn$ being a three-dimensional integer.
 Now consider the two-particle system in this finite box
 and let us take the center-of-mass frame of the system
 so that the two particles have opposite three-momentum
 $\bk$ and $-\bk$ respectively.
 The exact energy of the two-particle
 system in this finite volume is denoted as: $E_{1\cdot2}(\bk)$.
 We now define a variable $\bar{\bk}^2$ via:
 \be
 \label{eq:two_particle_dispersion}
 E_{1\cdot 2}(\bk)=\sqrt{m^2_1+\bar{\bk}^2}
 +\sqrt{m^2_2+\bar{\bk}^2}\;.
 \ee
 Note that due to interaction between the two particles,
 the value of $\bar{\bk}^2$ differs from its free counter-part
 $\bk^2$ with $\bk$ being quantized according to
 Eq.~(\ref{eq:free_k}). It is also convenient to further
 define a variable $q^2$ as:
 \be
 q^2=\bar{\bk}^2L^2/(2\pi)^2\;.
 \ee
 which differs from $\bn^2$ due to the interaction between the two mesons.
 What L\"uscher's formula tells  us  is a direct relation
 of $q^2$ and the elastic scattering phase shift $\tan\delta(q)$
 in the infinite volume. In the simplest case of $s$-wave elastic scattering,
 it reads:~\cite{luscher91:finitea}
 \be
 \label{eq:luscher_cube}
 q\cot\delta_0(q)={1\over \pi^{3/2}}\calZ_{00}(1;q^2)\;,
 \ee
 where $\calZ_{00}(1;q^2)$ is the zeta-function which
 can be evaluated numerically once its argument $q^2$ is given.
 Therefore, if we could obtain the exact two-particle energy
 $E_{1\cdot 2}(\bk)$ from numerical simulations, we could
 infer the elastic scattering phase shift by applying L\"uscher's
 formula given above. Here we would like to point out that,
 the above relation is in fact only valid under certain assumptions.
 For example, the size of the box cannot be too small. In particular, it
 has to be large enough to accommodate free single-particle states.
 Therefore, in a practical simulation, one should check whether this
 is indeed realized in the simulation. Polarization effects are also
 neglected which are suppressed exponentially by $\calO(e^{-mL})$ where
 $m$ being the single-particle mass gap. Also neglected are mixtures
 from higher angular momenta.

 In the case of attractive interaction, the lowest two-particle
 energy level might be lower than the threshold which then
 renders the quantity $q^2<0$. The phase shift in the continuum, $\delta(q)$,
 is only defined for positive $q^2$, i.e. energies above the
 threshold. When $q^2<0$, it is related to yet another phase $\sigma(q)$ via:
 \be
 \tan\sigma_0(q)={\pi^{3/2}(-iq)\over\calZ_{00}(1;q^2)}\;,
 \ee
 where $(-iq) >0$ and the phase $\sigma_0(q)$ for pure imaginary $q$
 is obtained from $\delta_0(q)$ by analytic continuation:
 $\tan\sigma_0(q)=-i\tan\delta_0(q)$~\cite{luscher91:finitea,Sasaki:2006jn}.
 The phase $\sigma_0(q)$ for pure imaginary $q$ is of
 physical significance since if there exists a true
 bound state at that particular energy, we have $\cot\sigma_0(q)=-1$
 in the infinite volume and continuum limit.
 In the finite volume, however, the relation $\cot\sigma_0(q)=-1$ is
 modified to:~\cite{Sasaki:2006jn}
 \be
 \label{eq:bound_finite_volume_corrected}
 \cot\sigma_0(q)=-1+{6\over
 2\pi\sqrt{-q^2}}e^{-2\pi\sqrt{-q^2}}+\cdots\;,
 \ee
 where the finite-volume corrections are assumed to be small.
 Therefore, for $q^2<0$, we could compute $\tan\sigma(q)$ from
 Monte Carlo simulations and check the possibility of a
 bound state at that energy. Note that the quantity $\cot\sigma(q)$
 differs from its continuum value $(-1)$ by corrections
 that decay like $(1/p_{B}L)e^{-2\pi p_B L}$ with
 $p_B=2\pi\sqrt{-q^2}/L$ being the binding momentum.
 Therefore, if the state is loosely bound, i.e. $(-q^2)$ being positive
 but close to zero, the finite volume correction goes to zero very slowly.
 This makes this criterion rather difficult to apply directly.
 For example, in the case of the deuteron, the binding is only a few
 MeV resulting in a length scale that is prohibitively large for
 practical lattice volumes.

% \subsection{Twisted boundary conditions}

 In order to increase the resolution in momentum space, particularly close to the threshold,
 we have adopted the so-called twisted boundary conditions (TBC)~\cite{Bedaque:2004kc,Sachrajda:2004mi}
 for the valence quark fields. The strategy follows that in Ref.~\cite{Ozaki:2012ce}.
 Basically, the quark field $\psi_\btheta(\bx,t)$, when
 transported by an amount of $L$ along
 the spatial direction $i$ (designated by unit vector $\bfe_i$), $i=1,2,3$,
 will change a phase $e^{i\theta_i}$:
 \be
 \label{eq:twistBC}
 \psi_\btheta(\bx+L\bfe_i,t)=e^{i\theta_i}\psi_\btheta(\bx,t)\;,
 \ee
 where $\btheta=(\theta_1,\theta_2,\theta_3)$ is the twisted angle (vector) for
 the quark field in spatial directions.
 The conventional periodic boundary conditions corresponds to $\btheta=(0,0,0)$ and,
 without loss of generality, one can restrict to the case $0\le\theta_i\le \pi$
 for the twisting case.

 Note that one has the choice of the twisted angles for different flavors
 of quarks involved in the calculation. Strictly speaking,
 the same twisted angle vector $\btheta$
 should be applied to the valence and the sea
 quark fields. This is also referred to as the {\em full twisting} case
 which is a well-defined unitary approach.
 At the moment, however, all of the available
 gauge field configurations are generated without twisting, i.e. with
 $\theta_i=0$ for all quark flavors in the sea. Therefore, if we apply
 twisted boundary conditions only to a particular valence flavor, the theory is
 in principle not unitary. This is known as {\em partial twisting}.
 It has been shown recently that,
 in some cases, partial twisting is equivalent to full twisting~\cite{Agadjanov:2013kja}.
 In the other cases, however, the corrections due to partially twisted boundary conditions
 are shown to be exponentially suppressed if the size of the box is large~\cite{Sachrajda:2004mi}.
 We will assume that these corrections are small.
 \footnote{This makes sense since L\"uscher's formalism also requires that
 exponentially suppressed corrections are negligible.}
 In this calculations, we only twist the light quarks while the charm quark fields remain
 un-twisted.  This avoids possible problems due to annihilation diagrams
 in this process as suggested in Ref.~\cite{Agadjanov:2013kja}.

 If we introduce the new quark fields
 \be
 \psi'(\bx,t)=e^{-i\btheta\cdot\bx/L}\psi_\btheta(\bx,t)\;,
 \ee
 it is then easy to verify that $\psi'(\bx,t)$ satisfy the
 conventional periodic boundary conditions along all spatial
 directions: $\psi'(\bx+L\bfe_i, t)=\psi(\bx,t)$ for $i=1,2,3$ if
 the un-primed field $\psi_\btheta(\bx,t)$ satisfies the twisted boundary
 conditions~(\ref{eq:twistBC}). For Wilson-type fermions,
 this transformation is equivalent to the replacement of
 the gauge link:
 \be
 U_{\mu}(x)\Rightarrow U'_{\mu}(x)=e^{i\theta_\mu a/L} U_{\mu}(x)\;,
 \ee
 for $\mu=0,1,2,3$ and $\theta_\mu=(0,\btheta)$.
 In other words, each spatial gauge link is modified by a $U(1)$-phase.
 \footnote{Note that this indeed brings the new gauge field out of the $SU(3)$
 gauge group. However, since the practical implementation did not utilize
 the $SU(3)$ nature of the gauge field, this is not a problem.}

 Normal hadronic operators are constructed using the primed fields.
 For example, a quark bilinear operator
 $\calO_\Gamma(\bx,t)=\bar{\psi}'_f\Gamma\psi'_{f'}(\bx,t)$, after summing over
 the spatial index $\bx$, will carry a non-vanishing momenta: $\bp=(\btheta_f-\btheta_{f'})/L$.
 The allowed momenta on the lattice are thus modified to:
 \be
  \bk={2\pi\over L}\left(\bn + {\btheta\over 2\pi}\right)
 \ee
 where $\bn\in \mathbb{Z}^3$ is the three-dimensional integer, the same as in the case
 without twisted boundary conditions. By choosing different values of $\btheta$, we
 could obtain more values of $\bar{\bk}^2$, or $q^2$ that are substituted into
 the L\"uscher formula.

 Another issue that should be addressed in the case of twisted boundary conditions
 is the change of symmetries. It is known that the original L\"uscher formula in the $s$-wave
 has a nice feature that only $s$-wave scattering phase shift $\delta_0(k)$ enters the game.
 The next-order corrections come from $l=4$ $g$-wave contaminations which are usually quite small
 when the scattering close to the threshold is considered. This fact comes about due to the
 property of the cubic group. With twisted boundary conditions applied, however,
 the symmetry of the system is reduced to subgroups
 of the cubic group and the mixing of lower waves with the $s$-wave will generally show up.
 Note that, for generic values of $\btheta$, the symmetry of parity is even lost.
 Parity is a good symmetry only for special values $\theta_i=0$ or $\pi$.
 In these cases, the mixing of $p$-wave with
 $s$-wave in L\"uscher formula would not occur since parity is a good symmetry.
 To circumvent this problem, following Ref.~\cite{Ozaki:2012ce}, we have chosen to simulate
 both parity-conserving points with: $\btheta=(0,0,0)$, $\btheta=(0,0,\pi)$, $\btheta=(\pi,\pi,0)$,
 whose symmetry group being $O_h$, $D_{4h}$, $D_{2h}$, respectively
 and parity-mixing points with: $\btheta=(0,0,\pi/4)$ and $\btheta=(0,0,\pi/8)$
 whose symmetry group being $C_{4v}$. In the former case, L\"uscher formula
 is simply Eq.~(\ref{eq:luscher_cube}) if we neglecting higher partial waves.
 In the latter case, $s$-wave and $p$-wave will show up and the formula looks
 like
 \be
  \label{eq:luscher_mix}
 [q\cot\delta_0(q^2)-m_{00}][q^3\cot\delta_1(q^2)-m_{11}]=m^2_{01}\;,
 \ee
 where $m_{00}$, $m_{11}$ and $m_{01}$ are known functions (involving the so-called zeta functions) of $q^2$.

 \section{One- and two-particle operators and correlators}
 \label{sec:operators}

 Single-particle and two-particle energies are measured
 in Monte Carlo simulations by measuring corresponding correlation functions,
 which are constructed from appropriate interpolating operators with definite symmetries.

 \subsection{One- and two-particle operators for non-twisted case}

 Let us first construct the single meson operators for
 $D^\ast$ and $D$ whose quantum numbers $J^P$ being $1^-$ and $0^-$.
 For the pseudo-scalar charmed mesons,  we utilize the following local interpolating
 fields in real space:
 \be
 \label{eq:single_operators_defs}
 [D^+]:\ \calP^{(d)}(\bx,t) = [\bar{d}\gamma_5 c](\bx,t)\;,
 \ee
 together with the interpolating operator for its anti-particle ($D^-$):
 $\bar{\calP}^{(d)}(\bx,t)=[\bar{c}\gamma_5d](\bx,t)=[\calP^{(d)}(\bx,t)]^\dagger$.
 In the above equation, we have also indicated the quark flavor content of the operator
 in front of the definition inside the square bracket. So, for example, the operator in Eq.~(\ref{eq:single_operators_defs})
 will create a $D^+$ meson when acting on the QCD vacuum.
 Similarly, one defines $\calP^{(u)}$ and $\bar{\calP}^{(u)}$ with the
 quark fields $d(\bx,t)$ in Eq.~(\ref{eq:single_operators_defs}) replaced by $u(\bx,t)$.
 In an analogous manner, a set of operators $\calV^{(u/d)}_i$ are constructed
 for the vector charmed mesons $D^{*\pm}$ with the $\gamma_5$ in $\calP^{(u/d)}$ replaced
 by $\gamma_i$.
 A single-particle state with definite three-momentum $\bk$ is
 defined accordingly via Fourier transform, see e.g. Ref.~\cite{Meng:2009qt}:
 \be
 \label{eq:singlemeson_untwist}
  \calP^{(u/d)}(\bk,t)=\sum_\bx \calP^{(u/d)}(\bx,t)e^{-i \bk \cdot \bx}.
 \ee
 The conjugate of the above operator is:
 \be
  [\calP^{(u/d)}(\bk,t)]^\dagger=\sum_\bx [\calP^{(u/d)}(\bx,t)]^\dagger
  e^{+i \bk \cdot \bx}\equiv\bar{\calP}^{(u/d)}(-\bk,t).
 \ee
 Similar relations also hold for $\calV^{(u/d)}_i$ and $\bar{\calV}^{(u/d)}_i$.

 To form the two-particle operators, one has to consider the corresponding
 internal quantum numbers.
 Since the newly discovered $Z^\pm_c(3900)$ state is charged, showing that the
 isospin of the state is $I=1$. For the  $I^G(J^{PC})$ quantum numbers of interest
 and expressing in terms of particle contents explicitly, we have:
 \be
 1^+(1^{+c}):\ \left\{\begin{aligned}
 & D^{*+}\bar{D}^0+c\bar{D}^{*0}D^+
 \\
 & D^{*-}\bar{D}^0+c\bar{D}^{*0}D^-
 \\
 & [D^{*0}\bar{D}^0-D^{*+}D^-]
 +c[\bar{D}^{*0}D^0-D^{*-}D^+]
 \end{aligned} \right.
 \ee
 where $c=\pm 1$ corresponds to the charge parity of the neutral state $C(Z^0_c)=\mp$
 respectively~\cite{Liu:2008fh}.
 Since $Z^\pm_c(3900)$ was observed in $J/\psi\pi^\pm$ final states,
 according to $G$-parity, we expect that the combination with $c=+1$ to yield the signal
 for $Z_c(3900)$. Therefore, in terms of the operators
 defined in Eq.~(\ref{eq:single_operators_defs}), we have used
 \be
 \calV^{(d)}_i(\bk,t)\bar{\calP}^{(u)}(-\bk,t)
 +\bar{\calV}^{(u)}_i(\bk,t)\calP^{(d)}(-\bk,t)
 \;,
 \ee
 for a pair of mesons with back-to-back momentum $\bk$.
 In this paper, we refer to this system of two mesons as $(D\bar{D}^{*})^\pm$
 system.

 On the lattice, the rotational symmetry group $SO(3)$ is broken down
 to the corresponding point group.
 For the two-particle system formed by a $D^*$ and a $D$ meson,
 the quantum number $J^P$ of the two-particle system can only be $1^+$
 which transform according to $T_1$ of the cubic group.
 To avoid complicated Fierz rearrangement terms, we have put the
 two mesons on two neighboring time-slices. Thus, we use the following
 operator to create the two charmed meson state from the vacuum,
 \begin{widetext}
 \be
 \label{eq:two-particle-operator-nontwist}
 \calO^i_\alpha(t)=\sum_{R\in G}
 \left[\calV^{(d)}_i(R\circ\bk_\alpha,t+1)\bar{\calP}^{(u)}(-R\circ\bk_\alpha,t)
 +\bar{\calV}^{(u)}_i(R\circ\bk_\alpha,t+1)\calP^{(d)}(-R\circ\bk_\alpha,t)
 \right]
 \;,
 \ee
 \end{widetext}
 where $\bk_\alpha$ is a chosen three-momentum mode. The index
 $\alpha=1,\cdots,N$ with $N$ being the number of momentum modes
 considered in the calculation. In this particular case, we have $N=6$.
 In the above equation, $G=O(\mathbb{Z})$ designates the cubic group
 and $R\in G$ is an element of the group and we have used the
 notation $R\circ\bk_\alpha$ to represent the momentum obtained from
 $\bk_\alpha$ by applying the operation $R$ on $\bk_\alpha$.

 Note that in the above constructions, we have not included relative orbital
 angular momentum of the two particles, i.e. we are only studying
 the $s$-wave scattering of the two mesons. This is justified for this
 particular case since close to the threshold the scattering is always
 dominated by $s$-wave contributions.

 \subsection{One- and two-particle operators for the case of twisted boundary conditions}

% \textcolor{red}{Please give at least an example how these operators are constructed, for one particular irrep.}

% \textcolor{blue}{Did we check the one-particle dispersion relation in this case??
% If you look into Ref.~\cite{Ozaki:2012ce} you will find that this will give you some very close to zero
% momentum points, which will allow us to check the dispersion relation better,right?}

 We choose to apply the twisted boundary conditions on the up and the down quark fields
 while the charm quark fields remain un-twisted. The single-meson operators are constructed
 similar to Eq.~(\ref{eq:singlemeson_untwist}), using the primed fields for the
 up/down quark fields.  The only difference now is the discrete version of the rotational symmetry.
 It has been reduced from
 $O_h$ to one of its subgroups: $D_{4h}$, $D_{2h}$ or $C_{4v}$, depending on
 the particular choice of $\btheta$.
 The other structures (flavor, parity when applicable etc.) of the operators remain unchanged.
 The property of the pseudo-scalar operators $\calP^{\prime (u/d)}$ remains unchanged,
 the operators $\calV^{\prime(u/d)}_i$, however, which used to form a basis
 for the $T_1$ irrep of $O_h$ now have to be decomposed into new
 irreps of the corresponding subgroups:
 \be
 \label{eq:reduction}
 \begin{array}{ll}
 T_1 \mapsto A_2 \oplus E                & D_{4h}\\
 T_1 \mapsto B_1 \oplus B_2 \oplus B_3   & D_{2h} \\
 T_1 \mapsto A_1 \oplus E                & C_{4v}
 \end{array}
 \ee
 Take the first line of Eq.~(\ref{eq:reduction}), for example, which corresponds to
 the case of $\btheta=(0,0,\pi)$,
 the original operator triplet $(\calV^{\prime(u/d)}_1,\calV^{\prime(u/d)}_2,\calV^{\prime(u/d)}_3)$
 should be decomposed into a singlet $\calV^{\prime(u/d)}_3$ and a doublet $(\calV^{\prime(u/d)}_1,\calV^{\prime(u/d)}_2)$
 which forms the basis for $A_2$ and $E$ irreps, respectively.

 The construction of the two-particle operators in the case of twisted boundary conditions
 is analogous. Taking the case of $\btheta=(0,0,\pi)$ as an example,
 the corresponding operators are
 \begin{widetext}
 \ba
 \label{eq:two-particle-operator-twist}
 \calO^{(A_2)}_\alpha(t) &=&\sum_{R\in G}
 \left[\calV^{\prime(d)}_3(R\circ\bk_\alpha,t+1)\bar{\calP}^{\prime(u)}(-R\circ\bk_\alpha,t)
 +\bar{\calV}^{\prime(u)}_3(R\circ\bk_\alpha,t+1)\calP^{\prime(d)}(-R\circ\bk_\alpha,t)
 \right]
 \;,
 \\
 \calO^{(E)}_{i,\alpha}(t)&=&\sum_{R\in G}
 \left[\calV^{\prime(d)}_i(R\circ\bk_\alpha,t+1)\bar{\calP}^{\prime(u)}(-R\circ\bk_\alpha,t)
 +\bar{\calV}^{\prime(u)}_i(R\circ\bk_\alpha,t+1)\calP^{\prime(d)}(-R\circ\bk_\alpha,t)
 \right]
 \;.
 \ea
 \end{widetext}
 The two-particle operators for the other cases are constructed similarly.

 \subsection{Correlation functions}

 One-particle correlation function, with a definite three-momentum $\bk$,
 for the vector and pseudo-scalar charmed mesons are defined respectively as,
 \ba
 \label{eq:single-particle-correlators}
  C^\calV(t,\bk) &=& \langle \calV^{(u/d)}_i(\bk,t)\bar{\calV}^{(u/d)}_i(-\bk,0)\rangle\;,
  \nonumber \\
  C^\calP(t,\bk) &=& \langle \calP^{(u/d)}(\bk,t)\bar{\calP}^{(u/d)}(-\bk,0)\rangle
  \;.
 \ea
 From these correlation functions, it is straightforward to obtain the single
 particle energies $E_{D}(\bk)$ and $E_{D^*}(\bk)$ for various lattice momenta $\bk$.

 We now turn to more complicated two-particle correlation
 functions. Generally speaking, we need to evaluate a (hermitian) correlation
 matrix of the form:
 \be
 \label{eq:correlation}
  C_{\alpha\beta}(t)=\langle \calO^{i\dagger}_{\alpha}(t) \calO^i_{\beta}(0) \rangle,
 \ee
 where $\calO^i_\alpha(t)$ represents the two-particle operator defined
 in Eq.~(\ref{eq:two-particle-operator-nontwist}).
 Similar correlation matrix is defined for the twisted case with operators
 properly replaced by its primed counterparts. Two particle energies that are to
 be substituted into L\"uscher's formula are obtained from this correlation
 matrix by solving the so-called generalized eigenvalue problem (GEVP):
 \footnote{We have used the matrix notation.}
 \be
 \label{eq:GEVP}
 C(t)\cdot v_\alpha(t,t_0)=\lambda_\alpha(t,t_0)C(t_0)\cdot v_\alpha(t,t_0)\;,
 \ee
 with $\alpha=1,2,\cdots,N$ and $t>t_0$.
 The eigenvalues $\lambda_\alpha(t,t_0)$ can be shown to behave like~\cite{luscher90:finite}
 \be
 \label{eq:exponential}
 \lambda_\alpha(t,t_0)\simeq e^{-E_\alpha(t-t_0)}+\cdots\;,
 \ee
 where $E_\alpha$ being the eigenvalue of the Hamiltonian for the system.
 This is the quantity we need from the simulation. This quantity,
 when converted into $q^2$, is then substituted into L\"uscher's formula
 for the extraction of the scattering information.
 The parameter $t_0$ is tunable and one could optimize the calculation
 by choosing $t_0$ such that the correlation function is more or less dominated
 by the desired eigenvalues at that particular $t_0$ (preferring a larger $t_0$) with
 an acceptable signal to noise ratio (preferring a smaller $t_0$).

 The eigenvectors $v_\alpha(t,t_0)$ are orthonormal with respect to the
 metric $C(t_0)$, $v^\dagger_\alpha C(t_0) v_\beta=\delta_{\alpha\beta}$ and
 they contain the information of the overlaps of the original operators with
 the eigenvectors. In fact, if we make a Cholesky decomposition of Hermitian
 matrix $C(t_0)=LL^\dagger$, the GEVP turns into an ordinary eigenvalue problem:
 \be
 L^{-1}C(t)L^{\dagger-1}\cdot (L^\dagger v)_\alpha
 =\lambda_\alpha(t,t_0)(L^\dagger v)_\alpha\;.
 \ee
 with the new eigenvectors: $u_\alpha=(L^\dagger v)_\alpha$. It is then easy to
 see that these eigenvectors form a $N\times N$ unitary matrix which
 transforms the original operators into the optimal linear combinations
 of operators that create the eigenstate of the Hamiltonian.

 Depending on different cases, we have chosen different number of
 two-particle operators in each symmetry sector.
 To be specific, for the non-twisted case,
 we have used $N=4$ in $T_1$, corresponding to $\bn^2=0,1,2,3$;
 for the twisted case of $\theta=\pi/4$ or $\theta=\pi/8$,
 we have used $N=3$ in $A_1$ and $E$, corresponding to $\bn^2=0,1,2$;
 for the case of $\btheta=(0,0,\pi)$ and $\btheta=(\pi,\pi,0)$
 we have used only $\bk_\alpha=(0,0,0)$ in each of the irreps.
 These information are listed in Table~\ref{tab:symmetries}.
\begin{table}
\centering \caption{Information about the two-particle operators used in
 this calculation together with the corresponding symmetries.
 Note that the last column lists the generic case of twisted BC
 for which we have taken $\theta=\pi/8, \pi/4$, respectively. The generic case
 distinguish itself from the rest since, in this case, parity is lost which
 causes $s$-wave $p$-wave mixing.} \label{tab:symmetries}
\begin{tabular}{|l|l|l|l|l|}
\hline
    &$\btheta=\bzero$    &$\btheta=(0,0,\pi)$ & $\btheta=(\pi,\pi,0)$
    & $\btheta=(0,0,\theta)$\\
\hline \hline
Symmetry  & $O_h$        & $D_{4h}$  & $D_{2h}$      & $C_{4v}$\\
\hline
irreps & $T_1$ & $A_2$, $E$ & $B_1$, $B_2$, $B_3$ & $A_1$, $E$ \\
\hline
Number of $\bk_\alpha$     &4     & 1,1  &1,1,1     &3,3\\
\hline
\end{tabular}
\end{table}

 Let us briefly comment on the multi-channel effects from $J/\psi\pi$ states.
 In principle, with the set of operators that we are using, which are $D-\bar{D}^*$ interpolating operators,
 do have certain overlap with the $J/\psi\pi$ states with the
 same quantum numbers. Note that this has nothing to do with the nature of the $Z_c(3900)$ state.
 Whatever nature it is, it couples to $J/\psi\pi$ and $D\bar{D}^*$ states simultaneously.
 Phenomenologically, the process $D+\bar{D}^*\rightarrow J/\psi \pi$ can
 be schematically viewed as a $D$ meson exchange, which should be small as long as
 the coupling is not outrageously large since
 the mass of the $D$ meson is rather heavy.
 Experimentally, there is also indications~\cite{Ablikim:2013xfr} that this mixing is small, namely
 $Z_c(3900)$ mainly couples to $DD^*$ states instead of $J/\psi\pi$ states although
 it was discovered in the $J/\psi\pi$ channel first.
 For the moment, we simply ignore this contribution and assume that a single-channel analysis is adequate.
 To really consider this multi-channel effect, one would need a coupled channel analysis
 involving both the $D\bar{D}^*$ operators and the $J/\psi\pi$ operators.
 What is more, one also needs the two-channel L\"uscher's formula instead of the
 single-channel L\"uscher formula~\cite{chuan05:2channel,Liu:2005kr,Lage:2009zv,Bernard:2010fp,Doring:2011vk,Doring:2012eu}.
 In that case, the $S$-matrix elements require $3$ parameters, all are functions of
 the energy.  A more sophisticated two-channel analysis involving both $D\bar{D}^*$ and $J/\psi\pi$
 operator is in progress and will be reported elsewhere~\cite{workinprogress}.

 \section{Simulation details and results}
 \label{sec:simulation_details}

 In this paper, we have utilized $N_f=2$ twisted mass gauge field configurations
 generated by European Twisted Mass Collaboration (ETMC) at $\beta=4.05$
 for three different pion mass values. Details of the relevant parameters are
 summarized in the following table.
\begin{table}
\centering \caption{Simulation parameters in this study. All lattices used
are of the size $32^3\times 64$ with lattice spacing $a\simeq 0.067$fm (or $\beta=4.05$).} \label{tab:parameter}
\begin{tabular}{|c|c|c|c|}
\hline
    &$\mu=0.008$    &$\mu=0.006$    &$\mu=0.003$\\
\hline \hline
$N_{\rm conf}$  &201                &214        &200\\
\hline
$m_\pi$[MeV]     &485             &420     &300\\
\hline
$m_\pi L$     &5.3              &4.6       &3.3\\
\hline
\end{tabular}
\end{table}

 For the valence charm quark, we have used the Osterwalder-Seiler action~\cite{Frezzotti:2004wz}.
 The up and down quark mass are fixed to the values of the sea-quark values
 while that for the charm quark is fixed using the mass of spin-averaged value of
 $J/\psi$ and $\eta_c$ on the lattice.
 The relevant quark propagators, in both single-meson and two-meson correlation functions
 discussed in the previous section, are computed using the corresponding wall sources
 without any smearing of the gauge links, for details see e.g. Ref.~\cite{Meng:2009qt}.
% %\begin{widetext}
% \begin{figure}[htb]
%  {\resizebox{0.45\textwidth}{!}{\rotatebox{-90}{\includegraphics{Dstar_mu006_pmodeall_ratio_plateau.eps}}}}
%  {\resizebox{0.45\textwidth}{!}{\rotatebox{-90}{\includegraphics{D_mu006_pmodeall_ratio_plateau.eps}}}}
% \caption{Effective mass plateaus for the single $D^*$ (upper panel) and $D$ (lower panel) mesons at %$\mu=0.006$.
% \label{fig:single_particle_plateaus}}%
% \end{figure}
% \end{widetext}
% In Fig.~\ref{fig:single_particle_plateaus}, we have shown the effective mass plateaus
% for these single particle energies
% from which the single particle dispersion relations are also verified.

\begin{figure}[htb]
   {\resizebox{0.45\textwidth}{!}{\includegraphics{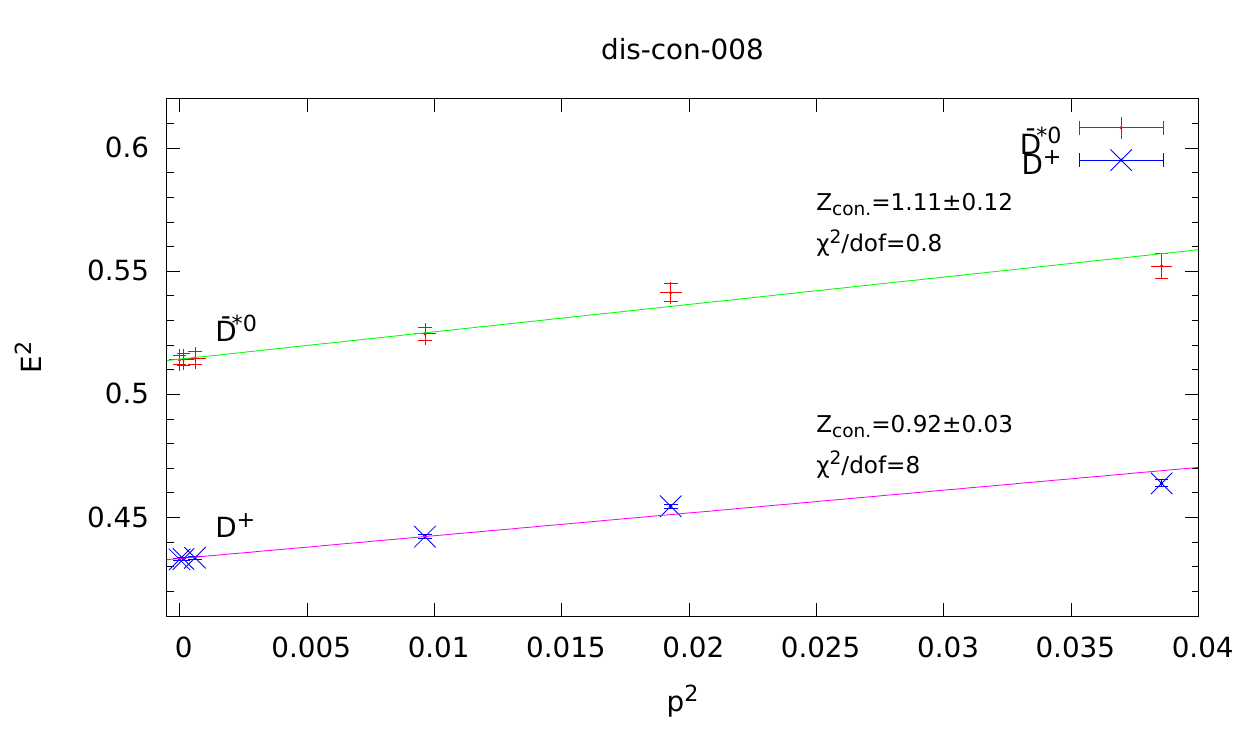}}}
   {\resizebox{0.45\textwidth}{!}{\includegraphics{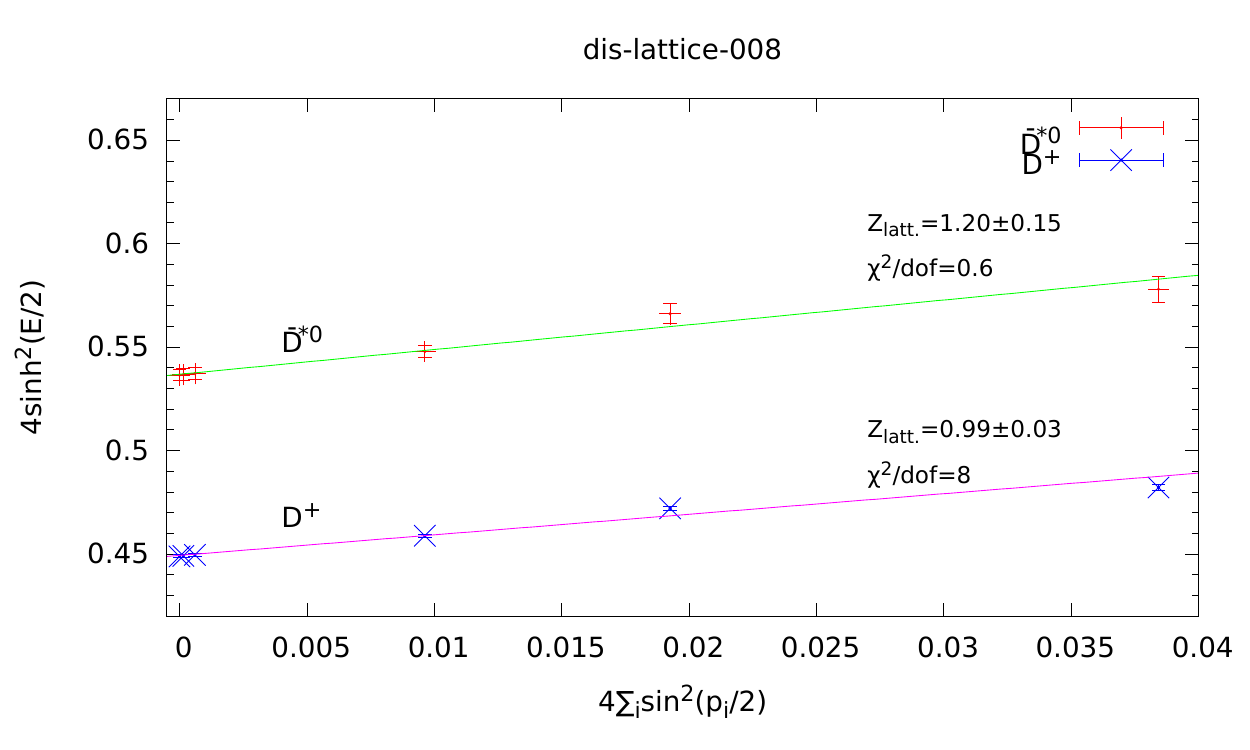}}}
  \caption{Dispersion relation for the $\bar{D}^{*0}$ (upper lines and data in each panel)
  and $D^+$ (lower lines and data in each panel) meson at $\mu=0.008$.
  The points with error bars are lattice data while the straight lines are the
  corresponding fits to the continuum (upper panel) and lattice (lower panel) dispersion relations.
  The values of $\chi^2/d.o.f$ for the fits are also shown in each panel.
  \label{fig:single_particle_dipersions}}%
  \end{figure}

 We have checked the single particle dispersion relations for the $D^\pm$ and $\bar{D}^{0*}$
 mesons, with both periodic boundary conditions and twisted boundary conditions.
 For the twisted boundary conditions, its equivalent small momentum points offer
 us a more stringent test for the dispersion close to zero momentum.
 We have performed fits for the dispersion relations for these mesons using
 both the usual continuum dispersion relation
 \be
 \label{eq:dis_continuum}
 E^2_\bp = m^2+Z_{\rm con.}\bp^2\;,
 \ee
 and its lattice counterpart
 \be
 \label{eq:dis_lattice}
 4\sinh^2{E_\bp\over 2} = 4\sinh^2{m \over 2}
 +Z_{\rm latt.}\sum^3_{i=1}4\sin^2{p_i\over 2}\;,
 \ee
 where $Z_{\rm con.}$ and $Z_{\rm latt.}$ being the corresponding
 speed of light squared parameter in the continuum and on the lattice,
 respectively. As we are interested only in the close to threshold scattering
 in this study, it suffices to check only the low momentum part of these dispersion
 relations where the difference of the two is negligible.
 This is indeed what we find for our charmed and anti-charmed mesons.
 The situation is illustrated in Fig.~\ref{fig:single_particle_dipersions}
 at $\mu=0.008$ for the $D^+$ and $\bar{D}^{*0}$ mesons.
 In this figure, we have taken only the six lowest momentum modes close to $\bp=\bzero$.
 The upper-panel in the figure corresponds
 to the continuum dispersion relation while the lower panel to that on the lattice.
 In each panel, the upper data and the straight line corresponds to $\bar{D}^{*0}$
 while the lower data and the straight line corresponds to $D^+$.
 The fitted values of $Z_{\rm con.}$, $Z_{\rm latt.}$ and the
 corresponding values for $\chi^2/d.o.f$ are also indicated.
 It is seen that the two dispersion relations yields
 compatible results which indicates that even for objects like
 charmed mesons, lattice artifacts are quite small. This is consistent with
 our previous experiences and might be due to one or several of the following
 reasons: the automatic $O(a)$ improvement of the twisted mass fermions,
 the smallness of our lattice spacing, and that we are studying low-energy
 scattering with small momenta.

 Apart from the single-particle dispersion relations, it has  also been suggested
 in previous lattice studies that it might be advantageous to also modify
 the two-particle dispersion relation in Eq.~(\ref{eq:two_particle_dispersion}), see e.g. Refs.~\cite{Lang:2011mn,Mohler:2012na}.
 We have also checked this possibility and found that, in our case, all $q^2$ values
 which eventually enter L\"uscher's formula (i.e. those values in Table~~\ref{tab:deltaE})
 are consistent within errors with those obtained from the continuum dispersion relation, i.e. Eq.~(\ref{eq:two_particle_dispersion}).
 We therefore simply take the values obtained from the continuum dispersion.

 \subsection{Extraction of two-particle energy levels}

 To extract the two-particle energy eigenvalues, we adopt
 the usual L\"uscher-Wolff method~\cite{luscher90:finite}.
 For this purpose,  a new matrix $\Omega(t,t_0)$ is defined as:
 \begin{eqnarray}
  \Omega(t,t_0)=C(t_0)^{-{1\over2}}C(t)C(t_0)^{-{1\over 2}},
 \end{eqnarray}
 where $t_0$ is a reference time-slice. Normally one picks a
 $t_0$ such that the signal is good and stable.
 The energy eigenvalues for the two-particle system are
 then obtained by diagonalizing the matrix $\Omega(t,t_0)$.
 The eigenvalues of the matrix have the usual exponential decay behavior
 as described by Eq.~(\ref{eq:exponential}) and therefore
 the exact energy $E_\alpha$ can be extracted from
 the effective mass plateau of the eigenvalue $\lambda_\alpha$.

 The real signal for the eigenvalue in our simulation turns out
 to be somewhat noisy. To enhance the signal, the following ratio was attempted:
 \begin{eqnarray}
  \calR_\alpha(t,t_0)={\lambda_\alpha(t,t_0)\over C^{\calV}(t-t_0,\bzero)C^{\calP}(t-t_0,\bzero)}
  \propto e^{-\Delta E_\alpha\cdot (t-t_0)}\;,
 \end{eqnarray}
 where $C^{\calV}(t-t_0,\bzero)$ and $C^{\calP}(t-t_0,\bzero)$ are one-particle correlation
 functions with zero momentum for the corresponding mesons defined
 in Eq.~(\ref{eq:single-particle-correlators}).
 \footnote{Note however that, in the case of twisted boundary conditions,
 one-particle correlation functions $C^{\calV'}(t,\bzero)$ and $C^{\calP'}(t,\bzero)$
 do not really correspond to zero three-momenta when constructed using the
 primed operators. Therefore, we still divide the
 eigenvalues $\lambda_\alpha(t,t_0)$ by the one-particle correlation function
 in the non-twisted case, configuration by configuration.
 Thus, Eq.~(\ref{eq:DeltaE_def}) and Eq.~(\ref{eq:kbar}) are still valid.}
 Therefore, $\Delta E_\alpha$ is the difference of the two-particle
 energy measured from the threshold of the two mesons:
 \be
 \label{eq:DeltaE_def}
  \Delta E_\alpha=E_\alpha-m_{D^\ast}-m_{D}\;.
 \ee
 The energy difference $\Delta E_\alpha$ can
 be extracted from the plateau behavior of the effective mass function
 $\Delta E_{\rm eff}(t)$ constructed from the ratio $\calR_\alpha(t,t_0)$ as usual.
  For all of the fits, the resulting $\chi^2$ per degree of freedom is around or less than one
 and the range is searched for by minimizing the $\chi^2$ per degree of freedom.
 The final results for $\Delta E_\alpha$, together with the corresponding ranges from which the
 $\Delta E_\alpha$'s are obtained, are summarized in Table~\ref{tab:deltaE}.
 We only list the lowest two energy levels for the non-twisted case and the twisted cases of $\btheta=(0,0,\pi/8)$ and $\btheta=(0,0,\pi/4)$, since we are not going to use those higher energy levels to extract the scattering parameters in the following analysis.
 \begin{figure}[htb]
   {\resizebox{0.45\textwidth}{!}{\includegraphics{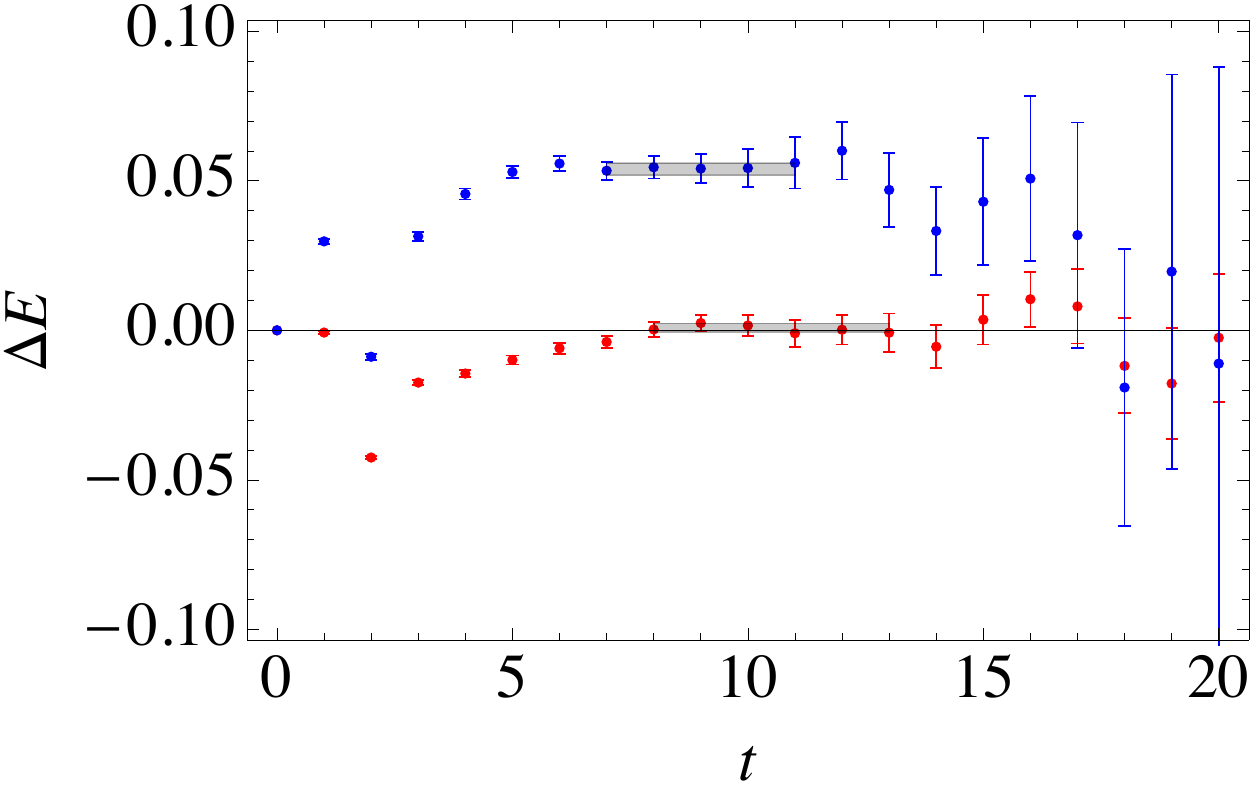}}}
   {\resizebox{0.45\textwidth}{!}{\includegraphics{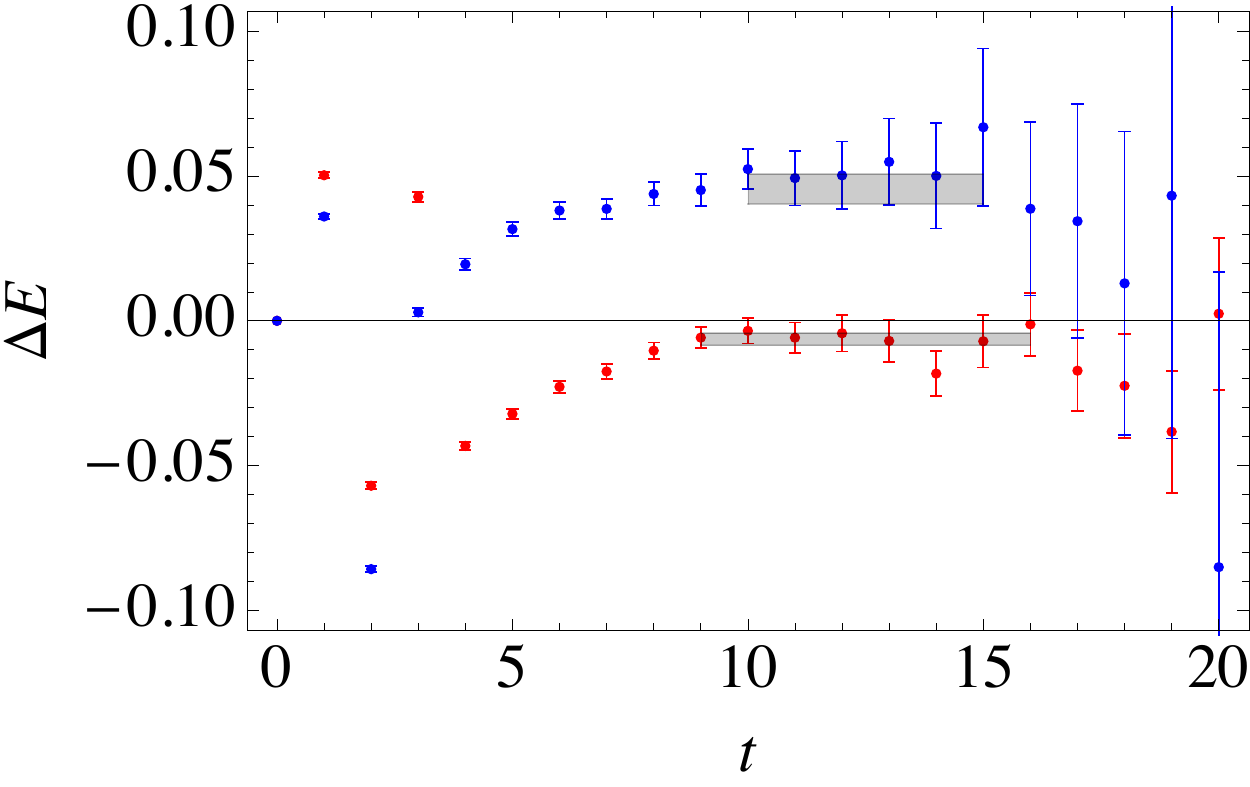}}}
   {\resizebox{0.45\textwidth}{!}{\includegraphics{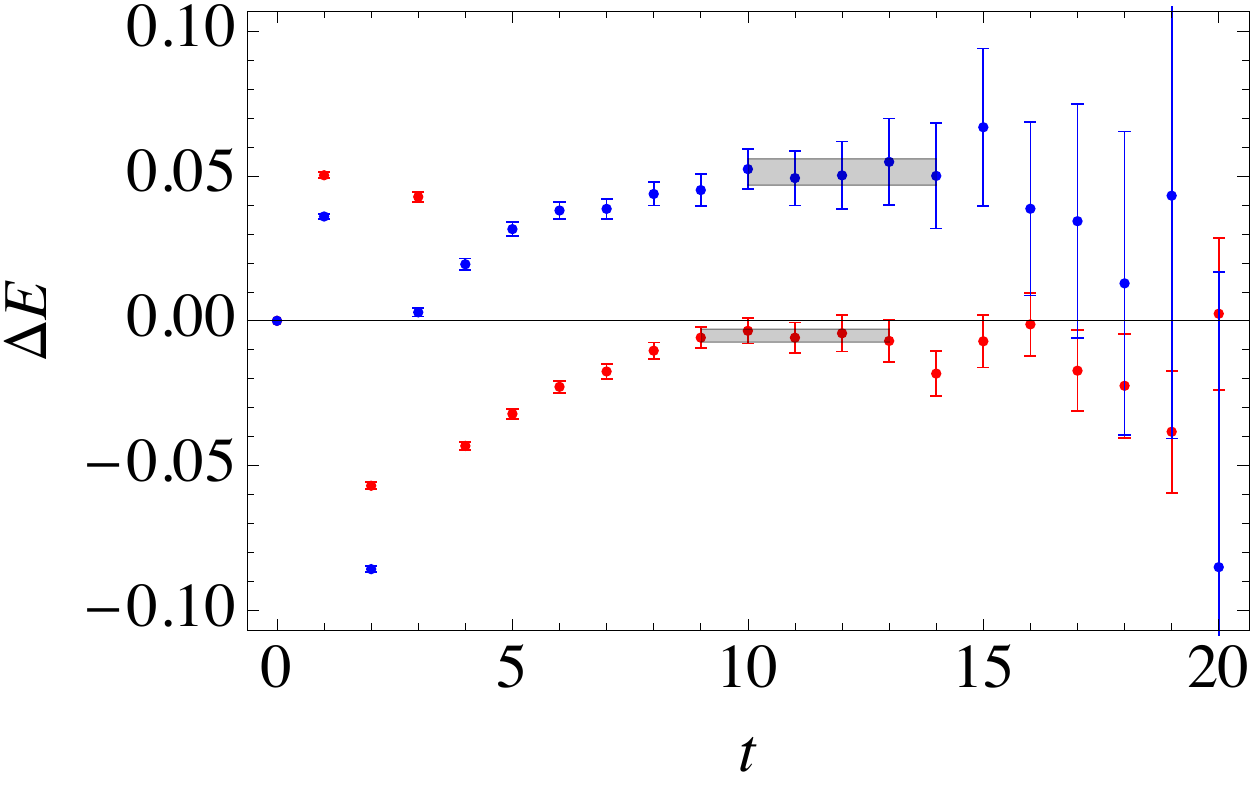}}}
%  {\resizebox{0.5\textwidth}{!}{\rotatebox{-90}{\includegraphics{z_3900_mu008_t_0_7_eigen0_5.eps}}}}
 \caption{Effective mass plots for the energy shift $\Delta E_\alpha$ at $\mu=0.003$ in
 the $A_1$ channel for $\btheta=\bzero$ (top), $\btheta=(0,0,\pi/8)$ (middle) and $\btheta=(0,0,\pi/4)$ (bottom).
 Two different colors indicate two lowest energy levels obtained from
  the variational calculation using $N=4$ (top) or $N=3$ (middle and bottom)
  different two-particle operators
  constructed using different $\bk_\alpha$ as described
  in Eq.~(\ref{eq:two-particle-operator-twist}). The grey horizontal bars indicate
  the fitted values for $\Delta E_\alpha$'s and the fitting ranges.
  \label{fig:two_particle_plateaus_A_1}}%
  \end{figure}
 As an illustration, in Fig.~\ref{fig:two_particle_plateaus_A_1},
 we have shown the effective mass plots and the fitted $\Delta E$'s at
 $\mu=0.003$ in the $A_1$ channel for three different values of $\btheta$:
 $\btheta=\bzero$, $\btheta=(0,0,\pi/8)$ and $\btheta=(0,0,\pi/4)$.
 In these cases, we have chosen $N=3$ different two-particle operators
 and only the two lowest  energy levels obtained from the (GEVP) process~(\ref{eq:GEVP})
 are shown using red and blue points.  Effective mass plots for other cases are similar.
 With the energy difference $\Delta E_\alpha$ extracted from the simulation data,
 one utilizes the definition:
 \be
   \label{eq:kbar}
  \sqrt{m_{D^\ast}^2+\bar{\bf k}^2}+\sqrt{m_{D_1}^2+\bar{\bf k}^2}
  =\Delta E_\alpha + m_{D^\ast} +m_{D}\;.
 \ee
 to solve for $\bar{\bk}^2\equiv (2\pi/L)^2q^2$ which is then plugged into L\"uscher's formula
 to obtain the information about the scattering phase shift.

\begin{table*}

%\begin{ruledtabular}
\begin{tabular}{|c|c||c|c||c|c||c|c|}
\hline
\hline
$\btheta$ & Irrep &  \multicolumn{2}{c|}{$\Delta E [t_{\min},t_{\max}](\mu=0.003)$}  &  \multicolumn{2}{c|}{$\Delta E [t_{\min},t_{\max}](\mu=0.006)$}  &  \multicolumn{2}{c|}{$\Delta E[t_{\min},t_{\max}] (\mu=0.008)$}\\
\hline
$\bzero$ & $T_1$ &0.001(1)[8,13] &0.054(2)[7,11] &-0.000(1)[10,14] &0.059(2)[7,11] &0.005(2)[13,17] &0.046(1)[7,11]\\
\hline
\multirow{2}{*}{$(0, 0, \frac{\pi}{8})$} &$A_1$ &-0.006(2)[9,16] &0.046(5)[10,15] &-0.005(2)[11,16] &0.051(2)[9,14] &0.005(4)[17,23] &0.056(4)[12,18]\\
%\cline{2-4}
&$E$ &0.005(2)[10,15] &0.061(2)[6,11] &0.016(5)[18,23] &0.064(2)[9,14] &-0.002(1)[4,10] &0.061(4)[14,20] \\
\hline
\multirow{2}{*}{$(0, 0, \frac{\pi}{4})$} &$A_1$ &-0.005(2)[9,13] &0.051(5)[10,14]&-0.004(2)[11,16] &0.052(2)[9,14] &0.006(2)[13,20] &0.056(4)[12,18] \\
%\cline{2-4}
&$E$ &0.005(2)[10,15] &0.061(2)[7,11] &0.022(8)[20,25] &0.065(2)[9,14] &-0.001(1)[4,12] &0.065(5)[14,20]\\
\hline
\multirow{2}{*}{$(0, 0, \pi)$} &$A_2$ &-0.015(5)[14,19] & &0.014(7)[19,24] & &0.021(5)[18,24] & \\
%\cline{2-4}
&$E$ &-0.003(10)[17,25] & &0.043(9)[20,27] & &0.028(6)[19,26] &  \\
\hline
\multirow{3}{*}{$(\pi, \pi, 0)$} &$B_1$ &0.003(10)[17,22] & &0.026(6)[18,26] & &0.059(8)[19,26] & \\
%\cline{2-4}
&$B_2$ &0.025(5)[12,17] & &0.031(1)[6,12] & &0.026(5)[16,22] & \\
%\cline{2-4}
&$B_3$ &0.029(1)[5,10] & &0.020(4)[14,21] & &0.029(1)[6,12] & \\
\hline
\hline
\end{tabular}
\caption{Results for the energy shifts $\Delta E$ obtained in our calculations for
various cases. The time interval $[t_{\min},t_{\max}]$ from which we extract the values of $\Delta E$ are
also listed. These ranges are relevant for the estimation of the error for the zeta functions as
described in the text.\label{tab:deltaE}}
%\caption{\label{Table:DeltaE} The values of energy shift $\Delta E$.}
%\end{ruledtabular}
\end{table*}

 \subsection{Extraction of scattering information}

 It is well-known that, close to the scattering threshold, the quantity $k\cot\delta(k)$
 has the following effective range expansion:
 \be
  {k^{2l+1}\cot\delta_l (k)}=a^{-1}_l+{1\over2} r_{l} k^2 +\cdots\;,
  \label{eq:kovertan}
 \ee
 where $a_{l}$ is the so-called scattering length, $r_{l}$ is the effective range
 for partial wave $l$ while $\cdots$ represents terms that are higher order in $k^2$.
 We will call $a_l$ and $r_l$ the low-energy scattering parameters in the following.
 It is more convenient to express this formulae in terms of $q^2$:
  \be
  {q^{2l+1}\cot\delta_l (q^2)}=B_l+{1\over2} R_{l} q^2 +\cdots\;,
  \label{qcotangent}
 \ee
 with $B_l=[L/(2\pi)]^{2l+1}a^{-1}_l$ and $R_l=[L/(2\pi)]^{2l-1}r_l$.
 Our task is to extract the parameters $B_l$ and $R_l$ from the simulation data.

 It is also well-known that, close to the threshold, scattering is dominated by phase shifts
 coming from lower partial waves as long as they are non-vanishing. We therefore will ignore
 all $l\ge 2$ partial waves in the L\"uscher formula for this study.
 Thus to extract these low-energy scattering parameters from the lattice data,
 we have to distinguish two different scenarios:
 the parity-conserving scenario, which corresponds to the non-twisting case
 and twisting case with special angles (i.e. those with $\theta=\pi$),
 and the parity-mixing scenario (those with values of $\theta\neq 0$ or $\pi$).
 Accordingly, the values of $q^2$ obtained are also categorized into
 two classes: the parity-conserving case and the parity-mixing case.
 The number of data points (i.e. number of $q^2$ values) in the two case is denoted as $N_0$ and $N_1$,
 respectively. So altogether we have $N_0+N_1$ points for $q^2$ values which
 are exactly those listed in Table~\ref{tab:deltaE}.

 The major difference between the parity-conserving data and parity-mixing data is as
 follows. As we have neglected all contributions from
 $l\ge 2$ partial waves, the parity-conserving data is only relevant
 for the $s$-wave scattering parameters $B_0$ and $R_0$ while parity-mixing data
 is relevant for both $s$-wave and $p$-wave scattering parameters: $B_0$, $R_0$, $B_1$ and $R_1$.
 In previous studies like Ref.~\cite{Ozaki:2012ce},
 the authors first used only the parity-conserving data to extract the
 $s$-wave scattering parameters. Then, the obtained scattering
 information for the $s$-wave is substituted into the fit for the $p$-wave parameters
 using the parity-mixing data. In this study, we attempt to simultaneously fit
 for all scattering parameters, both $s$-wave and $p$-wave, from all of our
 data points (both parity-conserving and parity-mixing).

 Just to make comparisons, we have attempted the following methods for
 the extraction of the scattering parameters:
 we could use only the parity-conserving data ($N_0$ data points) or
 all of our data ($N_0+N_1$ data points). In either of
 these cases, we could perform either the correlated fit or the uncorrelated fit.
 The detailed process will be described below with the correlated fit
 using all data as an example, which is more involved than other methods and yields the
 most reliable results. We regard these as our final results in this paper.
 However, for comparison purposes, the results for other cases
 are also tabulated for reference in Table~\ref{tab:result_all}
 and Table~\ref{tab:result_conserving}.

 To be specific, in the parity-conserving case, we define
 \be
 y_0(q^2)=q\cot\delta_0(q^2) \;.
 \ee
 According to L\"uscher's formula~(\ref{eq:luscher_cube}), this should be equal to
 \be
 m_{00}(q^2)={1\over \pi^{3/2}}\calZ_{00}(1;q^2)\;,
 \ee
 for the non-twisted case while for the twisted case of $\btheta=(0,0,\pi)$
 and $\btheta=(\pi,\pi,0)$ one simply replace the corresponding zeta function by
 $\calZ^{\btheta}_{00}(1;q^2)$.
 In the parity-mixing case, however, things are more complicated. Apart from the
 $s$-wave phase shift $\delta_0(q^2)$, L\"uscher formula
 will also involve $\delta_1(q^2)$ and maybe written as Eq.~(\ref{eq:luscher_mix}).
 We therefore define
 \be
 y_1(q^2)=\left[q\cot\delta_0-m_{00}\right]\left[q^3\cot\delta_1-m_{11}\right]
 \;.
 \ee
 which, according to L\"uscher formula, should be equal to $m^2_{01}(q^2)$.
 Note that in either case, the functions $m_{00}$, $m_{01}$, $m_{11}$ are
 all known functions of $q^2$ that involve various zeta-functions~\cite{Ozaki:2012ce}.
 In the following, these functions will be generally denoted as $Z(q^2)$ for convenience.
 In other words, $Z(q^2)$ stands for $m_{00}(q^2)$ and $m^2_{01}(q^2)$
 in the parity-conserving and parity-mixing case, respectively.

 One subtlety that concerns us is the estimation of errors for $Z(q^2)$ which
 are rapidly oscillating functions of $q^2$. These functions can also become divergent
 at specific values. The naive way of estimating the errors
 would be for each $q^2_I$ value and its error $\Delta q^2_I$, one simply substitutes
 $Z(q^2_I)$ for the central value and using $Z(q^2_I\pm \Delta q^2_I)$ for
 the estimation of the error. This is fine for some of our $q^2$ values but for
 $q^2$ values that are close to the divergent points of
 these functions, this results in extraordinarily large (and asymmetric) errors.
 We therefore attempted to estimate the errors for these functions
 directly from the data using the jack-knife method.

 To do this, recall that our values of $q^2_I$, with $I=1,\cdots,N_0+N_1$,
 are obtained from the corresponding energy shifts $\delta E_\alpha$
 as described in the previous section and then using Eq.~(\ref{eq:kbar})
 to convert into values of $\bar{\bk}^2$, or equivalently, $q^2$.
 In this process, we have obtained a set of jack-knifed,
 (Euclidean) time-dependent values for $q^2$:  $q^2_{I,a}(t)$,
 where $t$ denotes the time slice and $a$ indicates the corresponding value with
 the configuration numbered by $a$ left out.
 By searching an appropriate plateau in $t\in [t_{\min},t_{\max}]$,
 say by minimizing the $\chi^2$ per degree of freedom,
 we have obtained the values of $q^2_I$ using all of our configurations.
 These $q^2_I$ values are equivalent to the values of $\Delta E$
 listed in Table~\ref{tab:deltaE} with the help of Eq.~(\ref{eq:kbar}).
 The corresponding ranges $[t_{\min},t_{\max}]$ are also tabulated in Table~\ref{tab:deltaE}.
 Within the same temporal ranges that determine various values of $q^2_I$,
 we could define a (Euclidean) time-dependent zeta-function using
 the jack-knifed data sets $q^2_{I,a}(t)$ via
 \be
 Y^a_I(t)=Z(q^2_{I,a}(t))\;,\;\; t\in[t_{\min},t_{\max}]\;.
 \ee
 and also its average value:
 \be
 \bar{Y}_I(t)={1\over N}\sum^N_{a=1}Y^a_I(t)\;.
 \ee
 We then estimate the errors of $\bar{Y}_I(t)$ using conventional jackknife:
 \be
 \Delta Y_I(t)=\sqrt{{N-1\over N}\sum^N_{a=1}
 [Y^a_I(t)-\bar{Y}_I(t)]^2}\;.
 \ee
 In the next step, we define the weighted-average $Y^a_I$ over the temporal slices:
 \be
 \label{eq:weighted_average_Y}
 Y^a_I=\sum_{t} p_I(t) Y^a_I(t)\;,
 \ee
 with the probability $p(t)$ for time slice $t$ given by
 \be
 p_I(t)={[\Delta Y_I(t)]^{-2} \over \sum_t[\Delta Y_I(t)]^{-2}}
 \;,
 \ee
 where the summation is within the corresponding range of $[t_{\min},t_{\max}]$
 for that particular $q^2_I$. Note that the weighted average $Y^a_I$
 in Eq.~(\ref{eq:weighted_average_Y}) is
 equivalent to searching the plateau of $Y^a_I(t)$ in $t$, except that we
 demand that the range of this average should coincide with the range that
 we determined for the corresponding $q^2$ value.
 We can then define the expectation value
 \be
 \bar{Y}_I={1\over N}\sum_a Y^a_I\;,
 \ee
 and the corresponding covariance matrix,
 \be
 \label{eq:covariance}
 C_{IJ}={N-1\over N}\sum_a(Y^a_I-\bar{Y}_I)(Y^a_J-\bar{Y}_J)\;.
 \ee
 Thus, $C$ is an $(N_0+N_1)\times (N_0+N_1)$ matrix which incorporates also the
 correlations among $y_0$'s and $y_1$'s. This covariance matrix is estimated using
 our data sample and the corresponding inverse matrix $C^{-1}$ can also be obtained numerically.
 We stress that, though in some cases the matrix has rather large condition number (see below
 for further discussion), we had no practical problem in obtaining $C^{-1}$
 using the standard methods.

 For later convenience, we introduce an index function as follows,
 \be
 ind(I)=\left\{ \begin{aligned}
 0\;\; & \mbox{for $1\le I\le N_0$}\\
 1\;\; & \mbox{for $N_0+1\le I \le N_0+N_1$}
 \end{aligned}\right.
 \ee
 In other words, $ind(I)=0$ for the first $N_0$ parity-conserving data points
 while $ind(I)=1$ for the next $N_1$  parity-mixing data points. So our previous
 definitions of $y_0(q^2)$ and $y_1(q^2)$ may be written
 collectively as $y_{ind(I)}(q^2_I)$ with $I=1,2,\cdots, (N_0+N_1)$.

   \begin{figure}[htb]
   {\resizebox{0.45\textwidth}{!}{\includegraphics{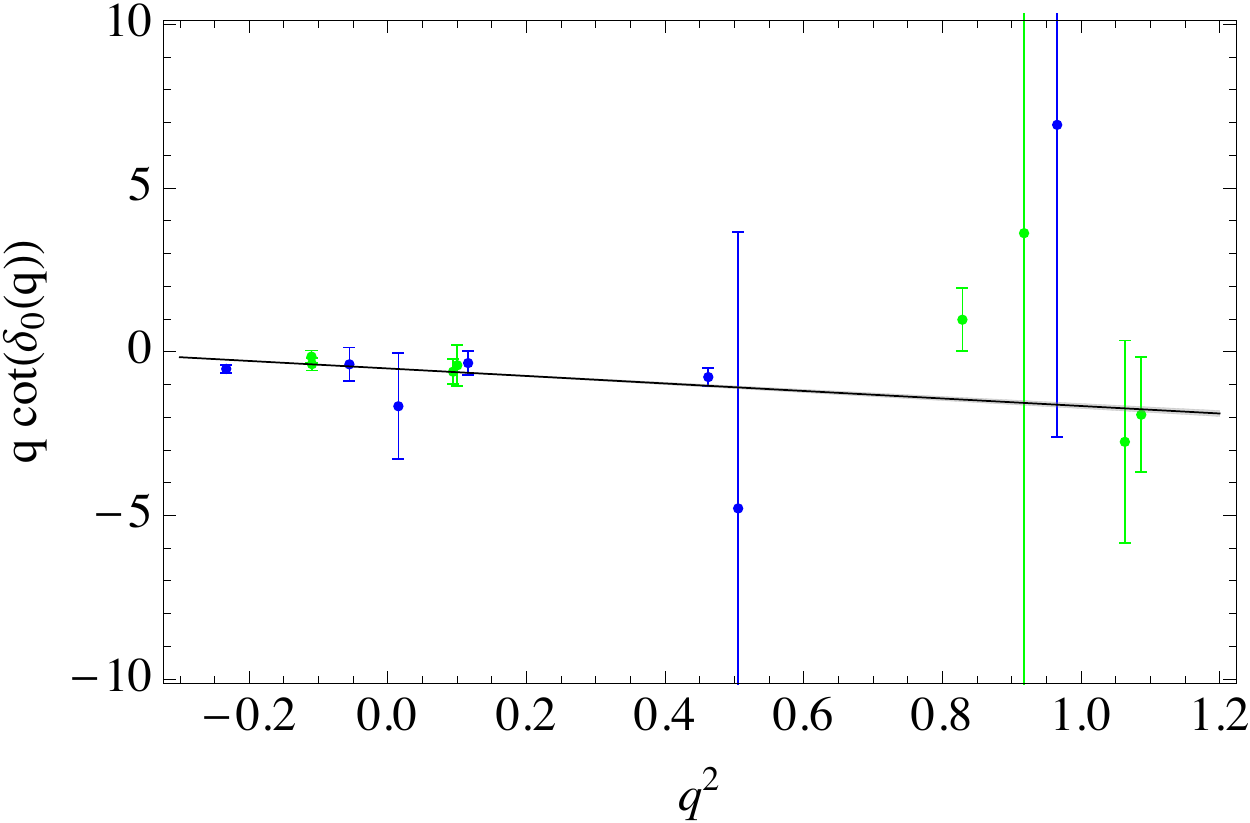}}}
    {\resizebox{0.45\textwidth}{!}{\includegraphics{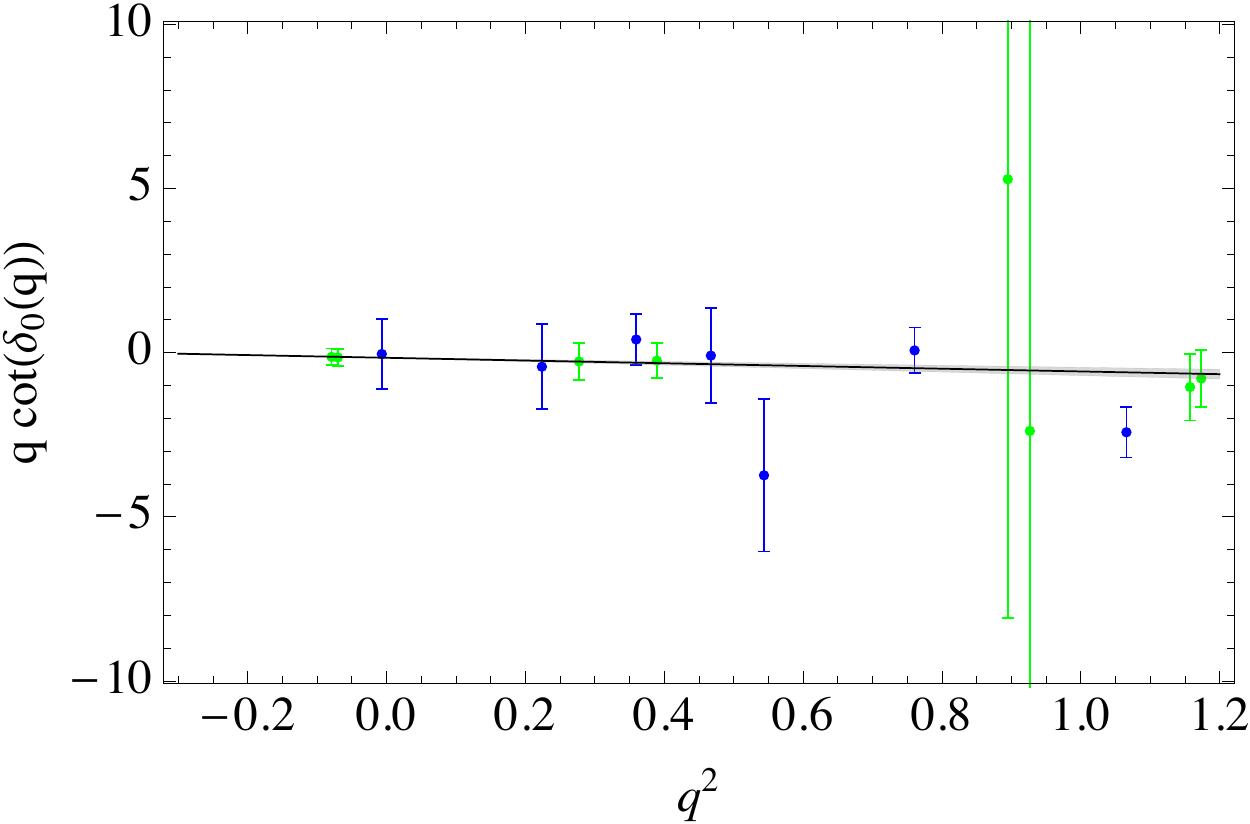}}}
     {\resizebox{0.45\textwidth}{!}{\includegraphics{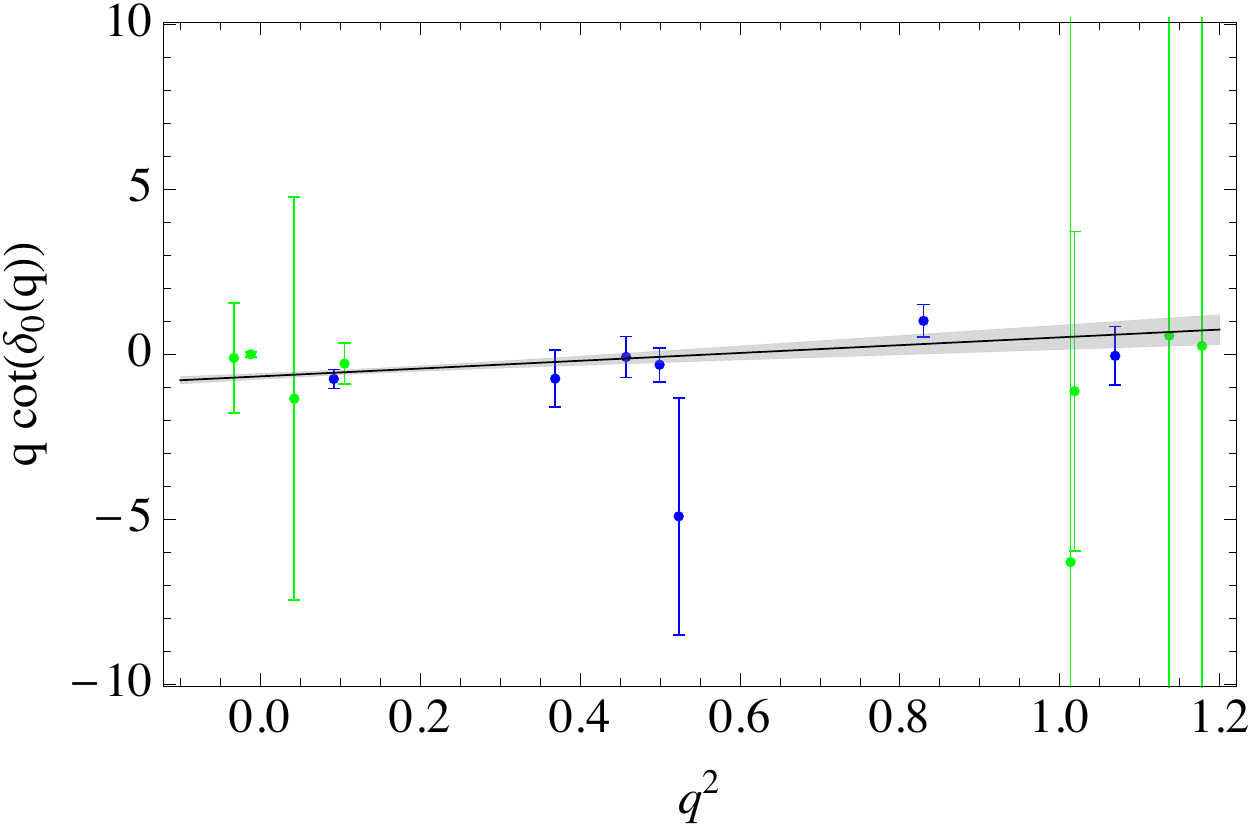}}}
 \caption{Results for the correlated fits as described in the text. Each panel, from
 top to bottom, corresponds to $\mu=0.003, 0.006$ and $0.008$, respectively.
 The quantity $q\cot\delta_0(q^2)$ is plotted versus $q^2$ for all our data points, both parity-conserving (blue)
 case and parity-mixing case (green). The straight lines and the bands indicate the fitted
 result for $F_0(q^2)=B_0+(R_0/2)q^2$ and the corresponding uncertainties in
 $B_0$ and $R_0$.\label{fig:swave_result}}%
  \end{figure}
 Finally, we can construct the $\chi^2$ function as usual
 \begin{widetext}
 \be
 \label{eq:target_chi2}
 \chi^2=\sum^{N_0+N_1}_{I,J=1}
 \left[F_{ind(I)}(q^2_I;\alpha)-y_{ind(I)}(q^2_I)\right]C^{-1}_{IJ}
 \left[F_{ind(J)}(q^2_J;\alpha)-y_{ind(J)}(q^2_J)\right]
 \;.
 \ee
 \end{widetext}
 where for $ind(I)=0,1$ the corresponding functions are
 (using the symbol $\alpha$ to collectively denote all
 the relevant parameters $B_0$, $R_0$, $B_1$ and $R_1$):
 \ba
 &&\!\!\!\!\!\!F_0(q^2;\alpha)=B_0+{1\over 2}R_0 q^2\;,\\
 &&\!\!\!\!\!\!F_1(q^2;\alpha)=[B_0+{R_0\over 2} q^2-m_{00}]
 [B_1+{R_1\over 2} q^2-m_{11}].
 \ea
 Minimizing the target $\chi^2$ function in Eq.~(\ref{eq:target_chi2}),
 one could obtain all the parameters, namely $B_0$, $R_0$, $B_1$ and $R_1$,
 in a single step with all of our data.
 This completes the process of correlated fit using all of our data.

 To get a feeling of the quality of the fits, we plot
 the quantity $q\cot\delta_0(q^2)$ vs. $q^2$ in Fig.~\ref{fig:swave_result}.
 This figure illustrates the situation for all three pion masses in our simulation.
 From top to bottom, each panel corresponds to $\mu=0.003$, $\mu=0.006$ and $\mu=0.008$, respectively.
 The data points obtained from our simulation are also plotted in these figures.
 The blue points are the data points from the parity-conserving case
 while the green points are the data for the parity-mixing case.
 For the former case, the errors for the data points are estimated using jack-knife method,
 i.e. the diagonal matrix element of the covariance matrix.
 In the latter case,  the values of $q\cot\delta_0(q^2)$ are obtained via
 the relation
 \be
 q\cot\delta_0=m_{00} +{m^2_{01}\over q^3\cot\delta_1(q^2)-m_{11}}
 \;,
 \ee
 where the quantity $q^3\cot\delta_1(q^2)$ on the r.h.s of the
 equation is replaced by $B_1+(R_1/2)q^2$ with the fitted values for $B_1$ and $R_1$.
 The errors for these points are estimated by the jack-knife method using the
 r.h.s. of the above equation.
 The straight lines and the grey shaded bands in the figure illustrates
 the function $F_0(q^2;\alpha)=B_0+(R_0/2)q^2$ and
 the uncertainties in the parameter ($B_0$ and $R_0$), respectively.
 As is seen from the figure, we get a reasonable fit for all three
 pion mass values.

   \begin{figure}[htb]
   {\resizebox{0.45\textwidth}{!}{\includegraphics{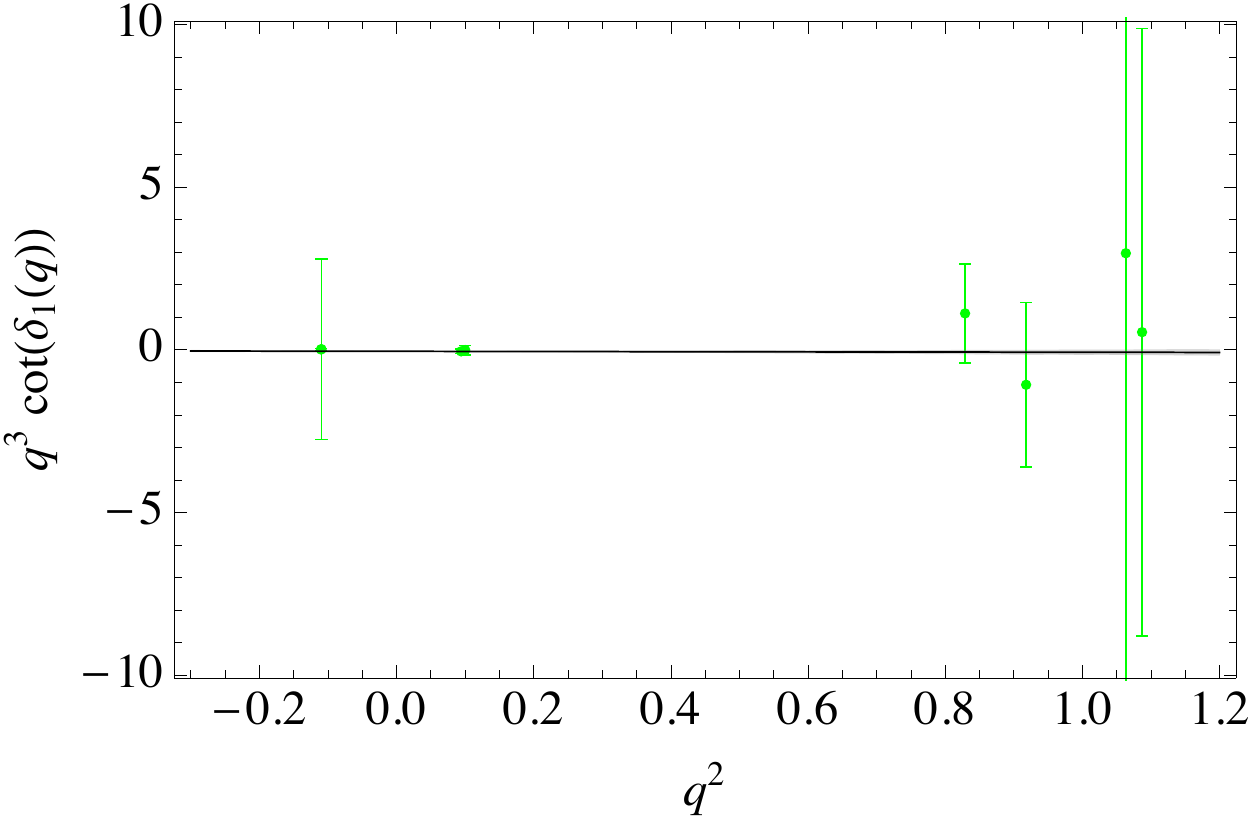}}}
 \caption{The quantity $q^3\cot\delta_1(q^2)$ vs. $q^2$ for the parity-mixing data at
 $\mu=0.003$.\label{fig:pwave_result}}%
  \end{figure}
 In a similar fashion, we could also plot the quantity
 $q^3\cot\delta_1(q^2)$ vs. $q^2$ for the parity-mixing data.
 This is shown in Fig.~\ref{fig:pwave_result} for $\mu=0.003$
 as an example.

 Note that, as far as the $s$-wave scattering parameters $B_0$ and $R_0$ are concerned,
 although they are most directly derived from the parity-conserving points
 (i.e. the blue points in Fig.~\ref{fig:swave_result}),
 the parity-mixing points (the green points in Fig.~\ref{fig:swave_result})
 also  help to  reduce the uncertainties in these parameters substantially.
 The effects coming from these points are folded in through the covariance
 matrix defined in Eq.~(\ref{eq:covariance}). To see this effect, one has to
 compare these results with the results obtained without the parity-mixing
 points. With the results listed in Table~\ref{tab:result_all} and
 Table~\ref{tab:result_conserving},
 it is seen that the parity-mixing points do indeed help to
 reduce the uncertainties in $B_0$ and $B_1$ in most cases.

\begin{table*}
%\begin{ruledtabular}
\begin{tabular}{|c||c|c|c|c|c|c|}
\hline
\hline
  & &$B_0$ &$R_0$ &$B_1$ &$R_1$ &$\chi^2/dof$ \\
  \hline
 \multirow{2}{*}{003} &Uncorrelated & -0.50(0.02) & -2.1(0.3)  &-0.02(0.01)  &-0.5(0.2)  &39.8/11\\
                                     & Correlated & -0.513(0.008) & -2.3(0.1) &-0.047(0.006) &-0.1(0.2) &47.0/11 \\
                                     & Correlated (omitted) & -0.35(0.12) & 0.8(0.6)  &-0.17(0.04) &1.00(0.09) &24.7/8\\
 \hline
\multirow{2}{*}{006}  &Uncorrelated & -0.176(0.005) & -1.1(0.1) & 0.4(0.1) &-3.1(0.5) & 15.8/11  \\
                                         & Correlated &-0.16(0.01) & -0.8(0.2) &0.29(0.05) &-2.6(0.3) &28.1/ 11\\
                                         & Correlated (omitted) &0.6(0.3) & -3.8(1.6) &-9.3(2.3) &17.8(5.0) &7.8/ 8\\
  \hline
\multirow{2}{*}{008}& Uncorrelated & -0.6(0.1) & 1.8(0.7)  &-0.02(0.01)  &0.4(0.5)  &9.6/11\\
                                        & Correlated & -0.67(0.09) & 2.4(0.8) &-0.037(0.008) &-0.1(0.2) &17.0/11 \\
                                        & Correlated (omitted) & -0.71(0.08) & 2.3(0.7) &0.02(0.03) &-0.2(0.2) &13.5/9 \\                                                                                  \hline
 \hline
\end{tabular}
\caption{Fit results with parity-conserving and parity-mixing points. \label{tab:result_all}}
%\end{ruledtabular}
\end{table*}

\begin{table}
%\begin{ruledtabular}
\begin{tabular}{|c||c|c|c|c|}
\hline
\hline
  & &$B_0$ &$R_0$  &$\chi^2/dof$ \\
  \hline
 \multirow{2}{*}{003} &Uncorrelated & -0.6(0.1) & -0.5(0.8)    &2.1/5\\
                                     & Correlated & -0.6(0.1) & -0.6(0.8)  &2.7/5 \\
 \hline
\multirow{2}{*}{006}  &Uncorrelated & 0.6(0.7) & -4.2(2.1)  & 6.4/5  \\
                                         & Correlated &1.0(0.7) & -4.5(1.8)  &6.5/ 5\\
  \hline
\multirow{2}{*}{008}& Uncorrelated & -0.9(0.3) & 3.4(1.2)    &4.6/5\\
                                        & Correlated & -0.8(0.3) & 3.8(1.1)  &5.5/5 \\
 \hline
 \hline
\end{tabular}
\caption{Fit results with parity-conserving data only.\label{tab:result_conserving} }
%\end{ruledtabular}
\end{table}

  \begin{figure}[htb]
   {\resizebox{0.45\textwidth}{!}{\includegraphics{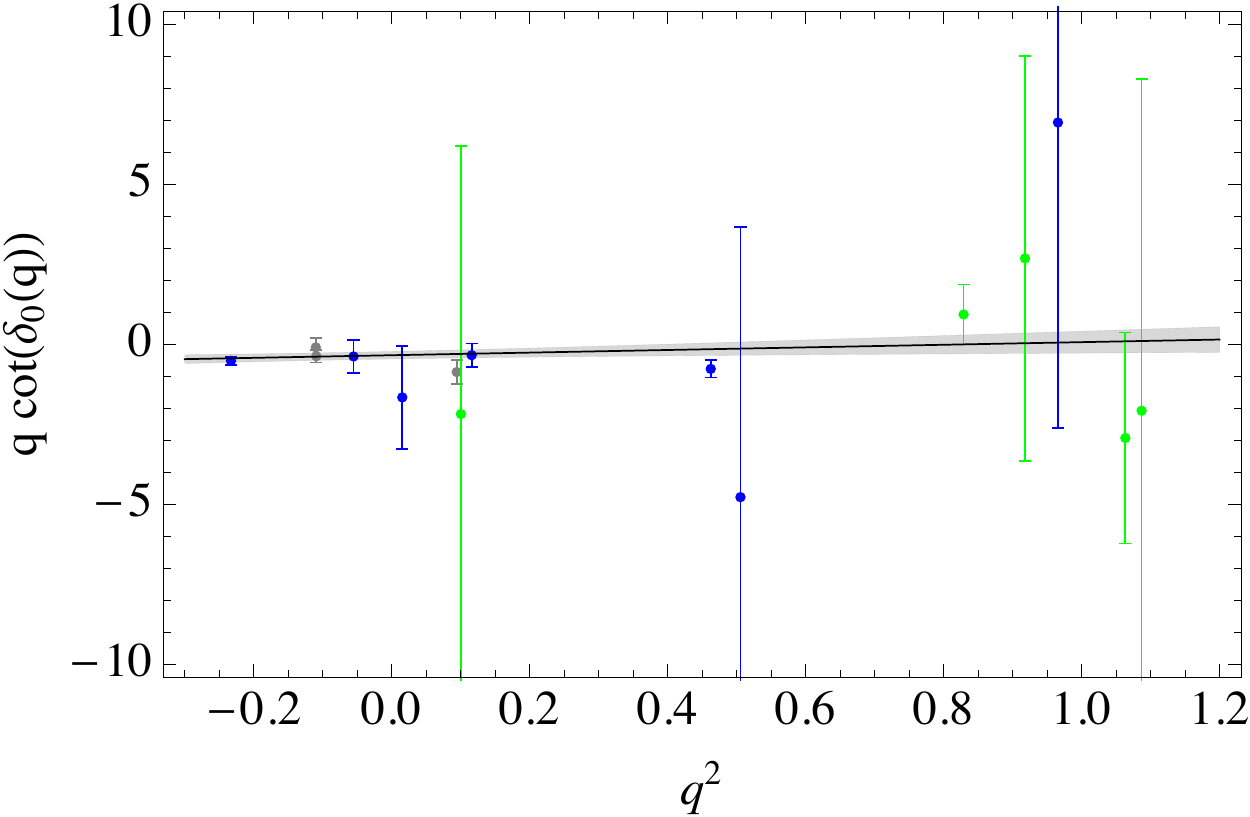}}}
    {\resizebox{0.45\textwidth}{!}{\includegraphics{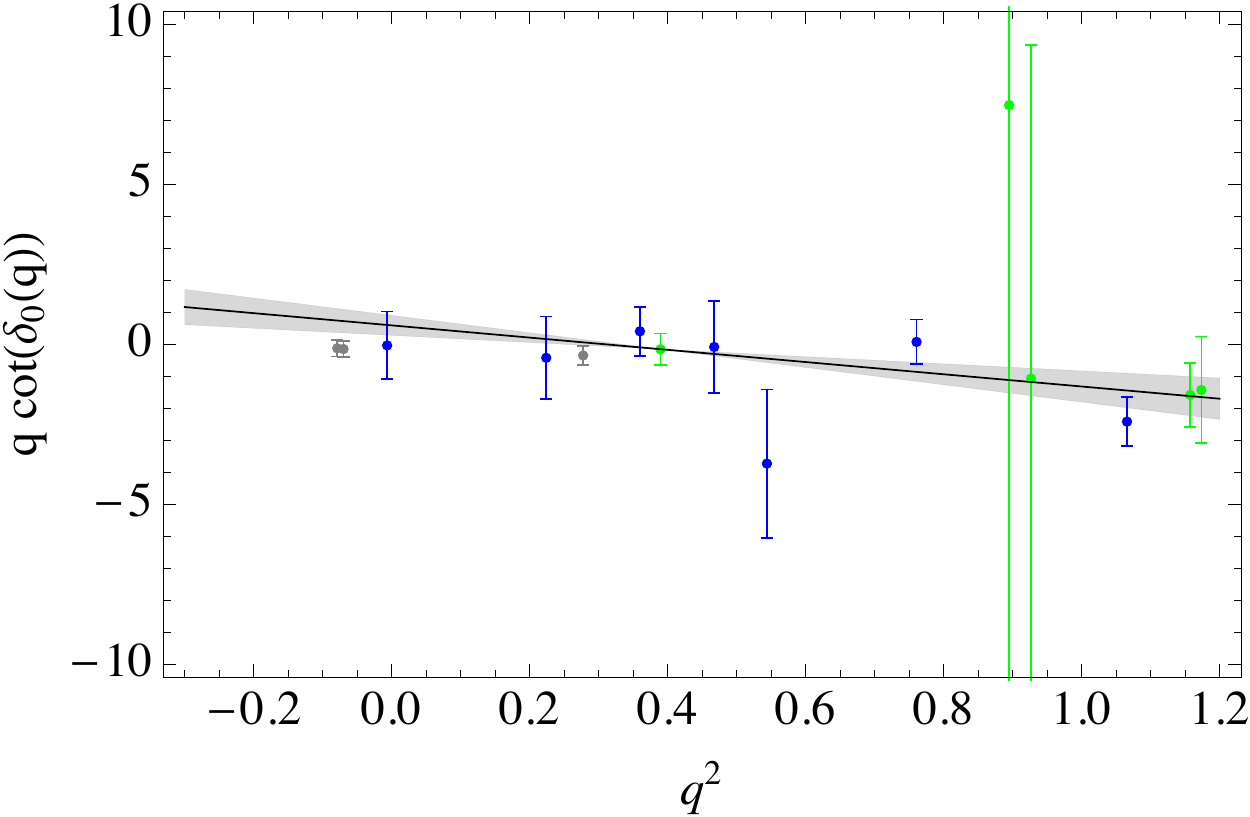}}}
     {\resizebox{0.45\textwidth}{!}{\includegraphics{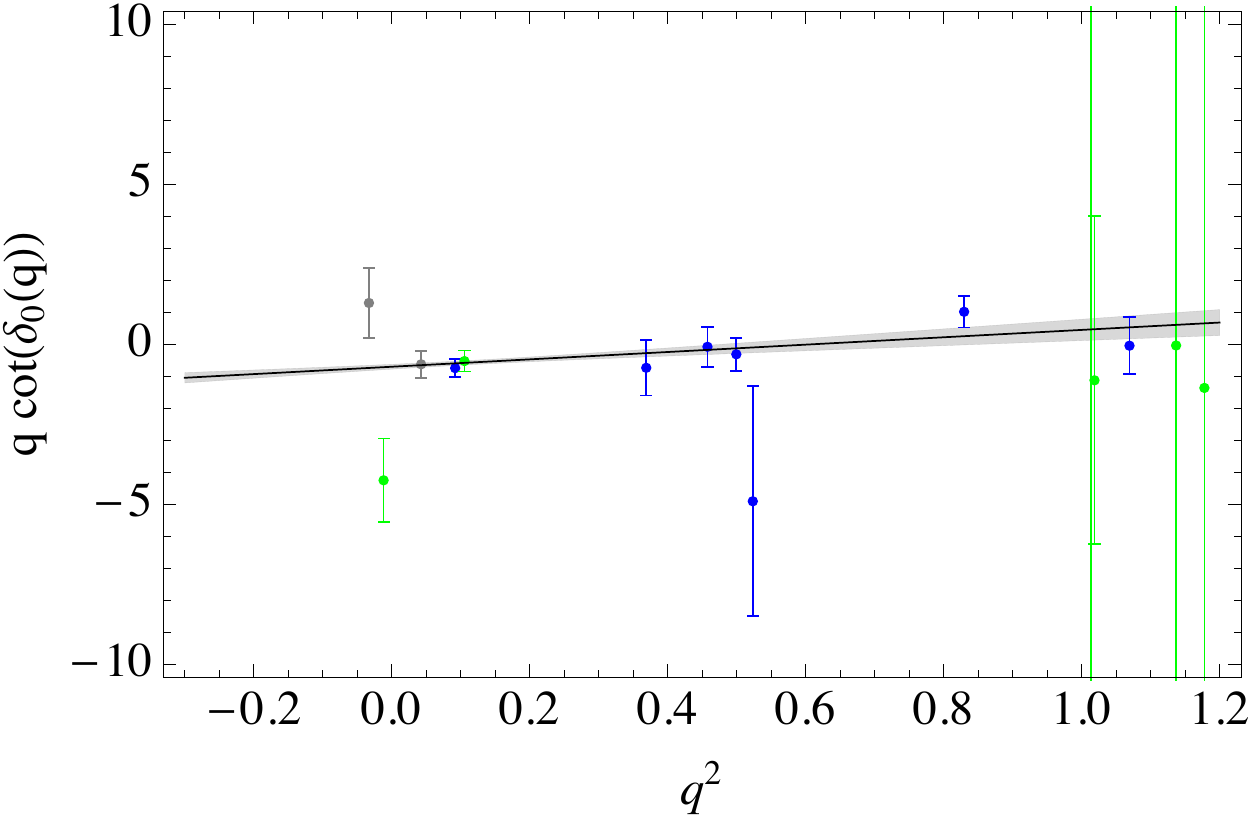}}}
 \caption{The same as Fig.~\ref{fig:swave_result} except that
 the grey data points are omitted in the $\chi^2$ fitting process for stability reasons
 as explained in the text. \label{fig:swave_result_omitted}}%
  \end{figure}
\begin{figure}[htb]
   {\resizebox{0.45\textwidth}{!}{\includegraphics{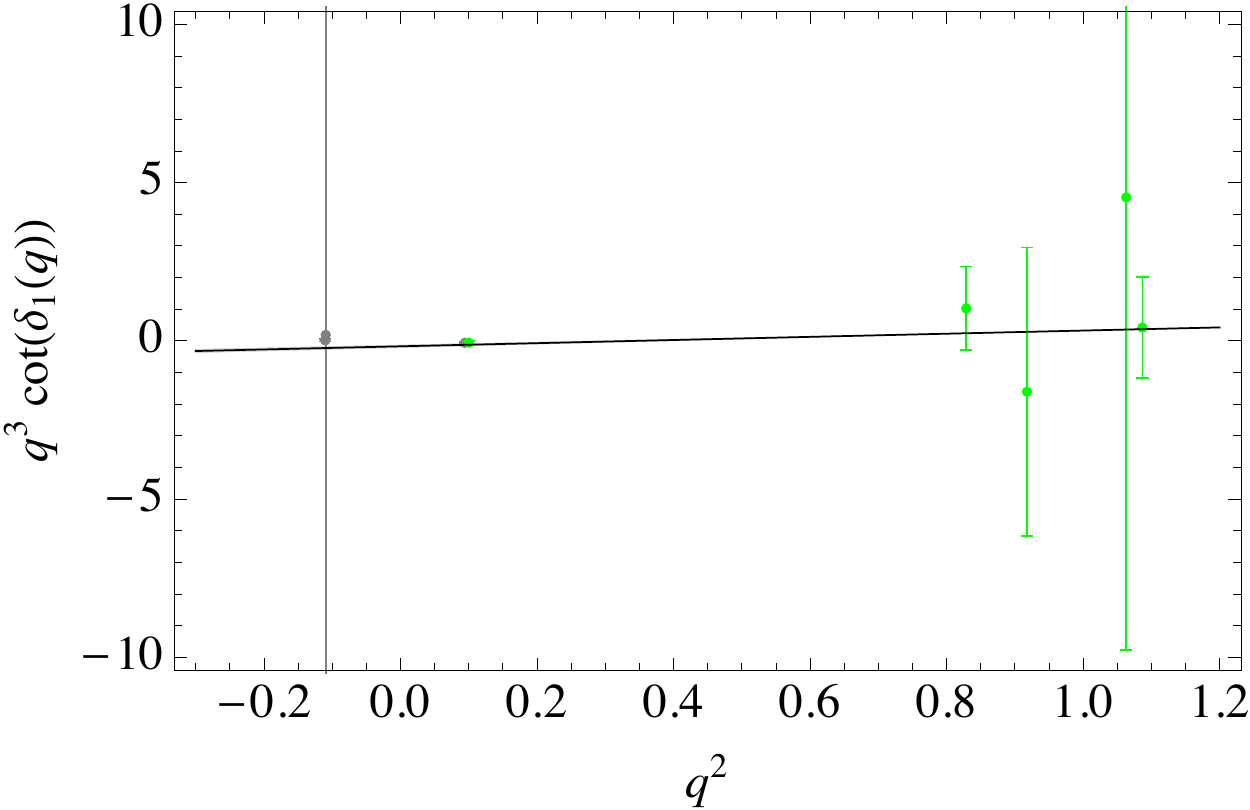}}}
 \caption{The same as Fig.~\ref{fig:pwave_result} except that the grey data points are omitted in the fit.\label{fig:pwave_result_omitted}}%
  \end{figure}

 In the course of inverting the covariance matrix $C$, it is found that in
 some cases the matrix is close to singular. This might bring up some potential worry
 about the stability of the fits. We studied this situation using the singular value decomposition
 method. We found that this close to singularity was caused by some of our $q^2$ values in some of the irreps
 in our calculation. To be specific, these correspond mainly to the lowest energy levels in irrep $A_1$ and $E$ at
 $\btheta=(0,0,\pi/8)$ from the parity-mixing data. Therefore,
 we have attempted the same fits as before except that with these data points omitted
 in the $\chi^2$ fitting process. This results in omitting $3$, $3$ and $2$ data points
 from $\mu=0.003$, $0.006$ and $0.008$, respectively. There is no well-established
 cut as to which points should be neglected in general but this procedure helps to
 give us some idea when compared with the results obtained with all the data.
 However, just to offer an idea where these omitted data points actually go, they are still plotted in
 the Fig.~\ref{fig:swave_result_omitted} and Fig.~\ref{fig:pwave_result_omitted}
 using grey data points. It is seen that the $B_0$ results for $\mu=0.003$ and $\mu=0.008$
 do not change much except that the errors are larger.
 For $\mu=0.006$, the central values of $B_0$ and $R_0$
  changed substantially with the corresponding errors are also much larger.
 For example, the estimate of $B_0$ changes from $-0.16(1)$ to $0.6(3)$, making
 the original value some $2.5\sigma$ below the new value.
  This is understandable from the middle panel in
 Fig.~\ref{fig:swave_result_omitted} where it is clearly seen that
 the three grey data points (the omitted ones) all lie significantly below the
 fitted straight line. This result is in fact in accordance with (consistent within errors) the
 results using only the parity-conserving data as listed in Table~\ref{tab:result_conserving}.
 Since there are no good reasons why these data points should be neglected in
 the first place and that they result in much larger errors, we think our original
 fits with all the data being more reasonable. However, the results with grey data
 points neglected are also tabulated for comparison.

 The fitted values for the scattering parameters are
 summarized  in Table~\ref{tab:result_all} for three values of $m^2_\pi$  in our simulation.
 As is said, we have performed both the correlated
 fits and uncorrelated fits. In the case of uncorrelated fits, we do not construct the covariance matrix
 as in Eq.~(\ref{eq:covariance}). We simply estimate the diagonal matrix
 elements $[\Delta Y_I]^{-2}$ using the conventional jackknife method.
 In the same table, under the title ``correlated (omitted)" for each parameter $\mu$,
 we have also listed the results with the grey data points omitted in the $\chi^2$ fitting process
 as explained above.
 Finally, we could also do our fits using only the parity-conserving data, as is
 done in previous studies~\cite{Ozaki:2012ce}. The results for the $s$-wave scattering
  parameters are listed in Table~\ref{tab:result_conserving} for comparison.
 As the correlation among different $Y_I$'s are quite substantial, especially
 those among $y_0$'s and $y_1$'s, as we observed from our covariance matrices,
 we regard our correlated fits with all of our data as being more reliable
 and they are taken as our final results.

 \subsection{Physical values for the scattering parameters}

 It is straightforward to convert the fitted values of $B_0$, $R_0$, $B_1$ and $R_1$ obtained
 in the previous subsection into physical units using the relation
 \be
 a_l=\left({L\over 2\pi}\right)^{2l+1}\left({1\over B_l}\right)\;,
 \;\;
 r_l=R_l\left({2\pi \over L}\right)^{2l-1}\;.
 \ee
 Then, if we take the numbers in Table~\ref{tab:result_all},
 we get for the $s$-wave scattering length $a_0$: $-0.67(1)$fm, $-2.13(13)$fm,
 $-0.51(7)$fm for $\mu=0.003$, $0.006$, $0.008$, respectively.
 The values for $r_0$ are also obtained accordingly.
 These numbers are summarized in Table.~\ref{tab:result_physical}

 It is  observed that the values we get for $a_0$ do not seem to
 follow a simple regular chiral extrapolation within the range that we have studied.
 We therefore kept the individual values for $a_0$ and $r_0$ for each case.
 This irregularity might be caused by the smallness of the value $m_\pi L\sim 3.3$ for
 $\mu=0.003$. To circumvent this, one has to study a larger lattice.

 The negative values of the parameter $B_0$ (hence the scattering length $a_0$) indicates
 that the two constituent mesons for the $(D\bar{D}^*)^\pm$ system
 have weak repulsive interactions at low energies.
 Therefore, our result does not support the bound state scenario for these
 two mesons. Recall that for an infinitely shallow bound state, we should have $B_0\sim 0^+$
 but our values of $B_0$ are all negative for all three pion mass values,
 as can be seen from Table~\ref{tab:result_all} and Table~\ref{tab:result_conserving}.
 The exceptions are the $\mu=0.006$ data sample using only the parity-conserving data
 or the correlated fit using all data but with three data points omitted.
 All these contradicts the possibility of a bound state,
 at least for the pion mass values we studied.
 \begin{table}
%\begin{ruledtabular}
\begin{tabular}{|c||c|c|c|}
\hline
\hline
  & $\mu=0.003$ &$\mu=0.006$  &$\mu=0.008$ \\
  \hline
  $a_0$[fm]& -0.67(1) & -2.1(1)    & -0.51(7)\\
 \hline
  $r_0$[fm]& -0.78(3) & -0.27(7)    & 0.82(27)\\
 \hline
\end{tabular}
\caption{The values for $a_0$ and $r_0$ in physical units obtained from
the numbers for the correlated fit in Table~\ref{tab:result_all}. \label{tab:result_physical}}
%\end{ruledtabular}
\end{table}

 Another check for the possible bound state would be
 to look for those negative $q^2$ values we obtained which corresponds to
 the negative values of $\delta E$ listed in Table~\ref{tab:deltaE}.
 However, one has to keep in mind that a negative value of $q^2$ does not
 necessarily signal a bound state in the infinite volume limit.
 Instead, for a finite volume,
 one has to check the condition in Eq.~(\ref{eq:bound_finite_volume_corrected}).
 The second term on the r.h.s of this equation indicates the size of the finite volume correction.
 This correction has to be small enough to justify the
 usage of this criterion since other higher order terms are neglected.
 We have checked all our data points with negative $q^2$
 and they do not seem to satisfy this condition. Therefore, our conclusion
 is that there is no indication of a bound state in this channel below the threshold, as
 far as we can tell from our data. This conclusion is consistent with a recent lattice study
 using Wilson fermions~\cite{Prelovsek:2013sxa}.
 Since the cases we are studying is still far from the physical pion mass case,
 we therefore still cannot rule out the possibility the appearance of a bound state once
 the pion mass is lowered (and the lattice size $L$ is also increased accordingly
 to control the finite volume corrections).
 Such scenarios do occur in lattice studies of two nucleons.

 \section{Conclusions}
 \label{sec:conclude}

 In this paper, we present an exploratory lattice study
 for the low-energy scattering of $(D\bar{D}^{*})^\pm$ two meson system near the threshold
 using single-channel L\"uscher's finite-size technique.
 The calculation was based on $N_f=2$ twisted mass fermion configurations of size $32^3\times 64$
 with a lattice spacing of about $0.067$fm.
 To investigate the pion mass dependence, three pion mass values are studied which
 corresponds to $m_\pi=300$MeV, 420MeV and 485MeV, respectively.
 To enhance the momentum resolution close to the threshold, twisted boundary conditions
 are also utilized together with the conventional periodic boundary conditions.
 Twisted boundary conditions also causes the mixing of $p$-wave with the $s$-wave
 scattering phase due to reduced symmetry. We have performed a combined analysis,
 using both the parity-conserving data and the parity-mixing data to
 obtain the scattering parameters.
 Our study mainly focuses on the $s$-wave scattering in the
 channel $J^P=1^+$ and the scattering threshold parameters, i.e. scattering length $a_0$ and effective range
 $r_0$ are obtained. An estimate for the $p$-wave scattering parameters
 are also obtained as a by-product.

 Our result indicates that the scattering lengths are negative,
 indicating a weak repulsive interaction between the the two mesons
 ($D$ and $\bar{D}^*$ or its conjugated systems
 under $C$-parity or $G$-parity). This is true for all three pion mass values that we simulated.
 We have also checked the possibility of the bound state for those negative energy shifts.
 None of those is consistent with a bound state. Our conclusion is that,
 based on our current lattice result,
 we do not support a bound state in this channel. Similar conclusion has been reached in
 a recent lattice study using Wilson fermions on a smaller lattice~\cite{Prelovsek:2013xba,Prelovsek:2013sxa}.
 However, as we pointed out already, we cannot rule out the possible appearance of
 a bound state for the two charmed mesons if the pion mass is lowered and the volume is
 increased accordingly.  This requires further more systematic lattice studies.
 Furthermore, it is also possible the the quantum numbers of the observed $Z_c(3900)$ is
 not $1^+$ or more complete set of interpolation operators and a coupled channel study is required.
 Thus, this lattice study has shed some light on the nature of $Z^\pm_c(3900)$ however
 it remains to be clarified by future studies.

 \section*{Acknowledgments}

 The authors would like to thank F.~K. Guo, U.~Meissner, A.~Rusetsky, C.~Urbach
 and B.~Knippschild for helpful discussions.
 The authors would also like to thank the European Twisted Mass Collaboration (ETMC)
 to allow us to use their gauge field configurations. Our thanks also go to
 National Supercomputing Center  in Tianjin (NSCC)
 and the Bejing Computing Center (BCC) where part of the numerical computations are performed.
 This work is supported in part by the
 National Science Foundation of China (NSFC) under the project
 No.11335001,
 No.11275169,
 No.11075167,
 No.11105153.
 It is also supported in part by the DFG and the NSFC (No.11261130311) through funds
 provided to the Sino-Germen CRC 110 ``Symmetries and the Emergence
 of Structure in QCD''.

% \bibliography{mybib}
% \bibliographystyle{apsrev4-1}


 \end{document}